\documentclass[sn-mathphys-num]{sn-jnl}

\usepackage{graphicx}%
\usepackage{multirow}%
\usepackage{amsmath,amssymb,amsfonts}%
\usepackage{amsthm}%
\usepackage{mathrsfs}%
\usepackage[title]{appendix}%
\usepackage{xcolor}%
\usepackage{textcomp}%
\usepackage{manyfoot}%
\usepackage{booktabs}%
\usepackage{algorithm}%
\usepackage{algorithmicx}%
\usepackage{algpseudocode}%
\usepackage{listings}%
\usepackage{natbib}
\usepackage[normalem]{ulem}

\usepackage{scalerel}
\usepackage{tikz}
\usetikzlibrary{svg.path}

\definecolor{orcidlogocol}{HTML}{A6CE39}
\tikzset{
  orcidlogo/.pic={
    \fill[orcidlogocol] svg{M256,128c0,70.7-57.3,128-128,128C57.3,256,0,198.7,0,128C0,57.3,57.3,0,128,0C198.7,0,256,57.3,256,128z};
    \fill[white] svg{M86.3,186.2H70.9V79.1h15.4v48.4V186.2z}
                 svg{M108.9,79.1h41.6c39.6,0,57,28.3,57,53.6c0,27.5-21.5,53.6-56.8,53.6h-41.8V79.1z M124.3,172.4h24.5c34.9,0,42.9-26.5,42.9-39.7c0-21.5-13.7-39.7-43.7-39.7h-23.7V172.4z}
                 svg{M88.7,56.8c0,5.5-4.5,10.1-10.1,10.1c-5.6,0-10.1-4.6-10.1-10.1c0-5.6,4.5-10.1,10.1-10.1C84.2,46.7,88.7,51.3,88.7,56.8z};
  }
}
\newcommand\orcidicon[1]{\href{https://orcid.org/#1}{\mbox{\scalerel*{
\begin{tikzpicture}[yscale=-1,transform shape]
\pic{orcidlogo};
\end{tikzpicture}
}{|}}}}

\usepackage{dcolumn}
\usepackage{bm}
\usepackage{hyperref}
\hypersetup{colorlinks=true, citecolor=blue, urlcolor=blue, linkcolor=blue}

\usepackage[mathlines]{lineno}

\theoremstyle{thmstyleone}%
\theoremstyle{thmstyletwo}%

\theoremstyle{thmstylethree}%

\begin{document}

\title[Fission of $^{180}$Hg and $^{264}$Fm: a comparative study]{Fission of $^{180}$Hg and $^{264}$Fm: a comparative study}


\author*[1,2]{R\'emi N. Bernard 
\orcidicon{0000-0002-2740-8984}}
 \email{remi.bernard@cea.fr }
\author[2,3]{C\'edric Simenel }
\author[4.5]{Guillaume Blanchon}
\author[2,3]{Ngee-Wein T. Lau}
\author[2,3,6]{Patrick McGlynn}
\affil*[1]{ CEA, DES, IRESNE, DER, SPRC, LEPh, 13115 Saint-Paul-l\`es-Durance, France}
\affil[2]{ 
Department of Fundamental and Theoretical Physics, Research School of Physics, Australian National University, Canberra, Australian Capital Territory 2601, Australia}
\affil[3]{ 
Department of Nuclear Physics and Accelerator Applications, Research School of Physics, Australian National University, Canberra, Australian Capital Territory 2601, Australia}
\affil[4]{
CEA,DAM,DIF, F-91297 Arpajon, France}
\affil[5]{Université Paris-Saclay, CEA, Laboratoire Matière sous Conditions Extrêmes, 91680 Bruyères-Le-Châtel, France}
\affil[6]{ 
Facility for Rare Isotope Beams, Michigan State University, East Lansing 48824, USA
}


\abstract{$^{180}$Hg is experimentally found to fission asymmetrically. This result was 
not expected as a naive fragment shell effects study would support the symmetric 
mode to be the most probable. In the present study we investigate both symmetric 
and asymmetric $^{180}$Hg fission modes at the mean field level using various 
multipole moment constraints. Potential energy surfaces are analysed in terms of 
shell effects that shape their topographies and connections to fragment shell effects 
are made. The non-occurrence of low energy symmetric fission is interpreted in terms 
of $^{90}$Zr fragment properties. Throughout this study a comparison with $^{264}$Fm 
and its symmetric doubly magic $^{132}$Sn fission fragments is done.}

\keywords{Nuclear fission, shell effects, mean field}



\maketitle

\section{Introduction}\label{intro}

From sub-lead systems to superheavy elements, the diverse fission behaviors of nuclei 
offer profound insights into the complex physics underlying the fission process. These 
various fission modes can often be interpreted through shell effects, which guide the 
compound nucleus from low deformation to scission. In the actinides, shell effects are 
predicted to manifest at various stages—near the saddle point, along the descent, or 
close to scission. 
Shell effects originating from the fragments 
frequently explain the nature of the most probable fission outcomes. In particular, 
spherical shell effects associated with magic numbers ($Z=50$ or $N=82$) have been 
invoked to explain asymmetric fission patterns in the actinide 
region, as illustrated by seminal works \cite{mayer1948,meitner1950,faissner1964,zhang2016,sadhukhan2016,CwiokPLB1994}. 
Deformed shell effects have later been shown to 
play a crucial role \cite{WilkinsPRC76}. In particular, octupole deformed  shell effects favor fragmentations into $Z\simeq 52-56$ or $N=88$ and have been proposed as a key driver fixing the final asymmetry in actinide fission \cite{ScampsNat18, ScampsPRC19}.

The potential for spherical shell gaps at $Z=40$ and $N=50$ to induce 
symmetric fission in the $^{180}$Hg isotope was naturally expected.
The mass yield measurements of $^{180}$Hg by Andreyev \textit{et al.} \cite{AndreyevPRL10}, 
conducted in 2010, presented surprising results that challenged our understanding of 
nuclear fission in the sub-lead region. The dominant asymmetric fission mode observed suggests 
that the expected spherical shell effects in the fragments are not the main 
drivers in this system. This  observation sparked immediate discussions, leading 
to contradictory interpretations. Initial analyses focused on Potential Energy Surface 
(PES) topographies, utilizing the Strutinsky correction method across various mean field 
approximations \cite{MollerPRC12,IchikawaPRC12,McdonnellPRC14}, where fragment shell 
effects were deemed negligible in the fission process. Conversely, other studies have 
highlighted the significance of deformed shell effects in fragments for explaining the 
asymmetric fission, employing scission point models \cite{andreev2012, andreev2013, PanebiancoPRC12}, 
mean-field calculations \cite{ScampsPRC19,WardaPRC12b}, and molecular structure arguments \cite{WardaPRC12b}.

In this context, $^{264}$Fm offers a compelling comparison to $^{180}$Hg, especially given 
its strong spherical shell effects akin to those in $^{90}$Zr. Hence, theoretical investigations 
into $^{264}$Fm's fission \cite{staszczak2009, sadhukhan2014, asano2004, MollerNPA87,simenel2014a, PascaEPJW18} 
predict a clear dominance of symmetric fission into $^{132}$Sn doubly magic fragments, underscoring the influence of spherical 
$Z~=~50$ and $N~=~82$ shell effects. This prediction, seemingly at odds with 
the $^{180}$Hg case, provides a unique lens through which to examine the puzzling fission 
behavior of $^{180}$Hg. 

The introduction of the smoothed level density (sld) method represents 
a recent advancement in analyzing the multitude of shell effects during the fission process \cite{BernardEPJA23}. 
Applied to any type of mean field approximations the sld method
focuses on the shell effects at the Fermi level of single particle spectra 
and highlights the way the compound nuclei find their way to fission
on a PES.
This study employs the sld method for both $^{180}$Hg and $^{264}$Fm across 
PES, investigating the role of low sld in guiding nuclei towards fission. 
The emergence of prefragments and their impacts are examined, with particular attention to 
the symmetric fission valley of $^{180}$Hg.
\\

\section{Framework}
\label{sec::framework}

We examine the quantities of interest within the Hartree-Fock-Bogoliubov (HFB) approximation, employing constraints alongside the Gogny D1S interaction, as detailed in previous studies 
\cite{BergerCPC91,RobledoJPG18,BernardPRC20}. Throughout our calculations, we maintain time reversal, simplex, and axial symmetries. PES are generated by constraining  
multipole moments among ($Q_{20}$, $Q_{30}$, $Q_{40}$), while allowing freedom for higher moments. These PES are constructed to represent the minimal HFB energy at each specified deformation, across a mesh of stepping through 2~b$^{l/2}$ in each $Q_{l0}$ multipole moment. For a coherent comparison of compound nuclei, we introduce the dimensionless $\beta_l$ variables, defined as 
\begin{equation}
\beta_{l} = \sqrt{\frac{2l+1}{4\pi}}
 \frac{4 \pi Q_{l0}}{3 r_{0}^l A^{(l/3+1)}}.
 \label{eq::beta}
\end{equation}
Here, $Q_{l0}$ is calculated as
\begin{eqnarray}
 Q_{l0} =\sqrt{\frac{4\pi}{2l+1}}\int d\vec{r} \rho(\vec{r}) Y_{l0}(\theta) |\vec{r}|^l,
 \label{eq::q}
\end{eqnarray}
with $Y_{l0}$ denoting the spherical harmonic functions, $\rho(\vec{r})$ the spatial density, 
and $r_0$ set to 1.2~fm. The harmonic oscillator basis, spanning the Fock space, comprises 14 major shells for $^{180}$Hg and its fragments, and 16 for $^{264}$Fm. Axial and radial oscillator lengths have been optimized for each HFB calculation within the gradient method framework \cite{egido1995, robledo2011a}. 
Along the 1D asymmetric paths from the spherical configuration to scission, they span from $(b_\perp,b_z)=(2.1,1.8)$~fm$^2$ to $(2.1,3.2)$~fm$^2$
for $^{180}$Hg and from $(b_\perp,b_z)=(1.9,2.8)$~fm$^2$ to $(2.1,3.5)$~fm$^2$ for $^{264}$Fm respectively. In both cases the radial length parameter stays in a small range while the axial one globally increases with the stretching of the nucleus.

A nucleus is deemed scissioned once the spatial density at its neck drops below $0.08$~fm$^{-3}$. The scission point and scission line are thus defined by the last HFB
state(s) before scission occurs in 1D and 2D respectively. 

Chemical potentials, serving as the Lagrange multipliers for particle number constraints, 
delineate the energies marking the Fermi levels within the nucleus. The Fermi gap for each 
isospin is the energy difference between the two closest single-particle states surrounding the 
chemical potential, with these states being chosen as eigenstates of the diagonal $h$ component of 
the HFB Hamiltonian matrix. In the absence of pairing, the chemical potential is positioned at 
the midpoint of the Fermi gap.

The smoothed level density $\eta$ at the Fermi level is evaluated within an energy window 
$[a,b]$ centered on the Fermi energy gap's midpoint $\epsilon_0=(a+b)/2$
\begin{eqnarray}
 \eta=\sum_{E_{i}\in [a,b]}f( E_{i}-\epsilon_0).
 \label{eq::eta}
\end{eqnarray}
Here, $E_i$ represents the energy of a single-particle state $i$ within the window $[a,b]$. 
As the nucleon
energy spectra are less compressed in lighter nuclei, for $^{180}$Hg, the energy window is set to 3.0~MeV, and for $^{264}$Fm, it is taken as 2.5~MeV. 
The smoothing function $f$ equals 1 at $E_{i}=\epsilon_0$ and decreases linearly to zero at the 
window's edges. This approach, favoring smoothed level densities over gaps, better accommodates 
intruder particle levels near the Fermi surface. The robustness of these results against the 
energy window's size has been verified in \cite{BernardEPJA23}.

Approaching scission, the minimal spatial density along the neck identifies the left 
and right prefragments. By integrating the spatial density to the left and 
right, we obtain the mass and charge numbers of the prefragments, along with their individual 
multipole moment deformations $Q_{l0}$ for $l\in [2,6]$.
The particle numbers are rounded to the closest even integer to highlight the shell effects at stake within the prefragments. It has been checked that the raw values are close to the rounded ones 
for both systems.
These particle numbers and geometric 
variables then serve as constraints for calculating the configurations of each prefragment.
\section{$\beta_{2}-\beta_{3}$ study of PES and SLD}
\label{sec::PESandgapA}

\begin{figure*}
  \centering
   \includegraphics[width=0.49\linewidth ]{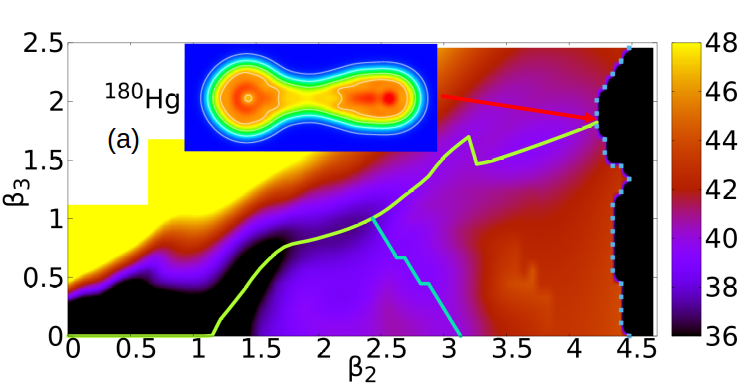}   
   \includegraphics[width=0.49\linewidth ]{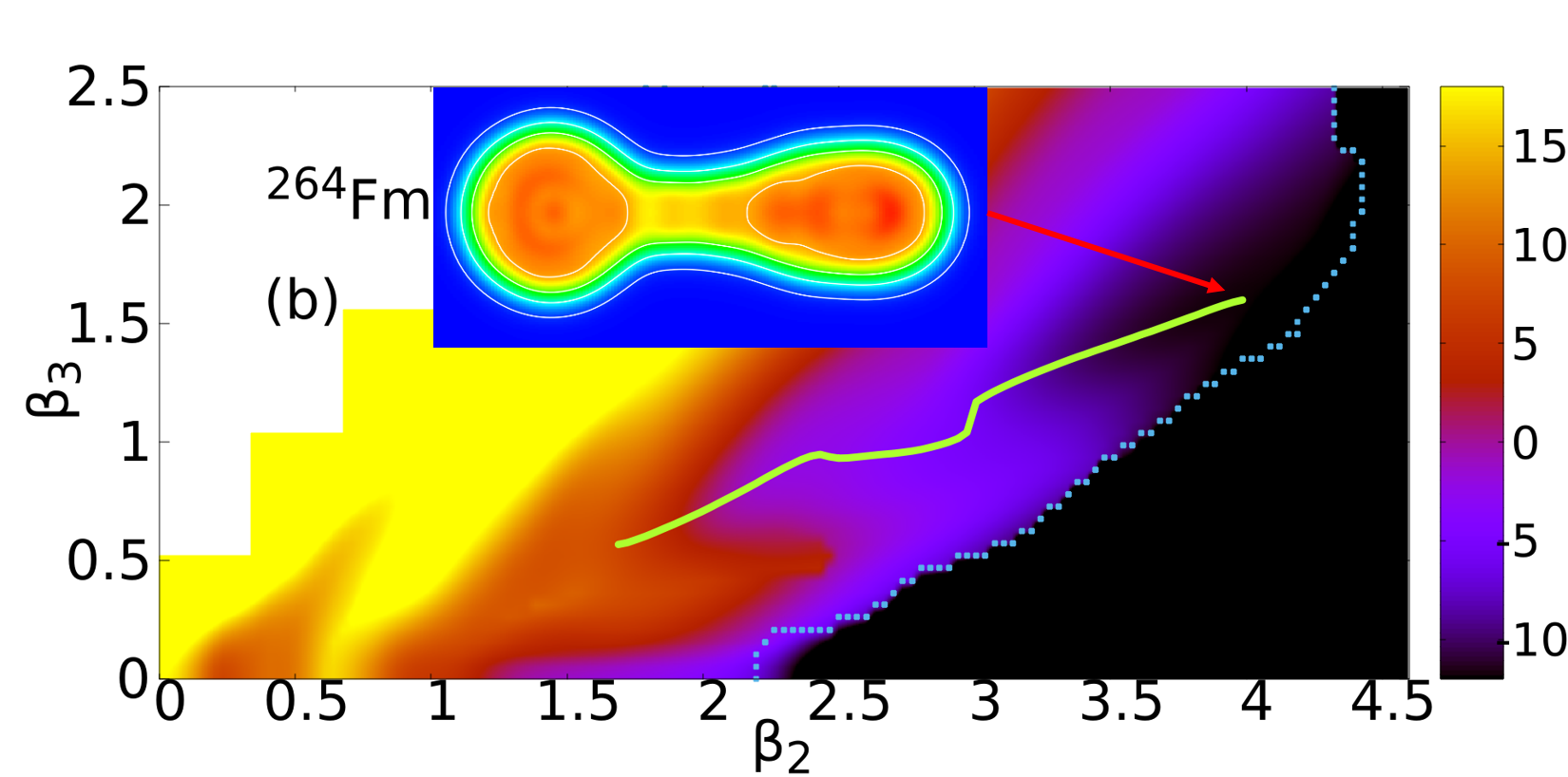} \\
   \includegraphics[width=0.49\linewidth ]{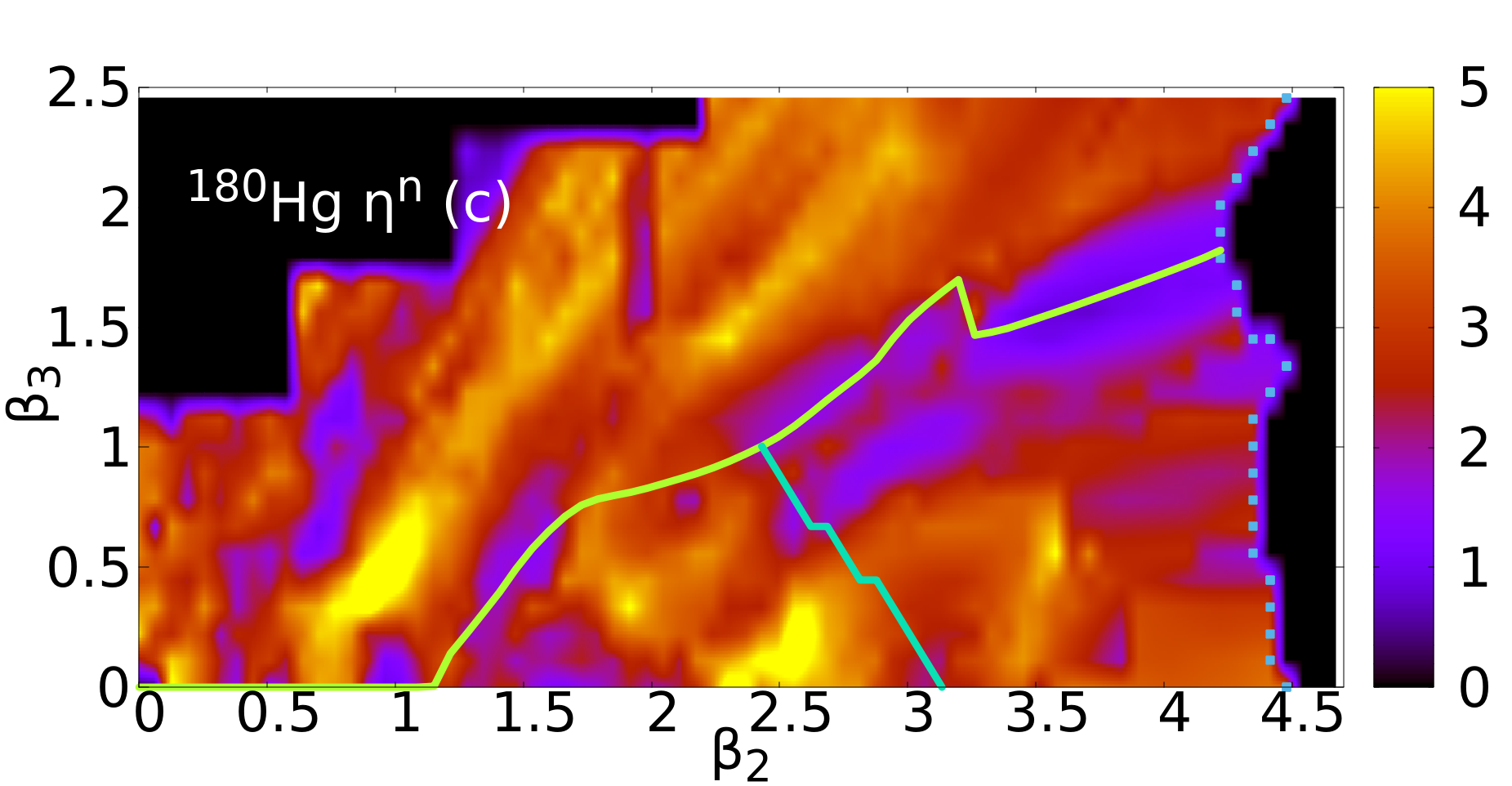} 
   \includegraphics[width=0.49\linewidth ]{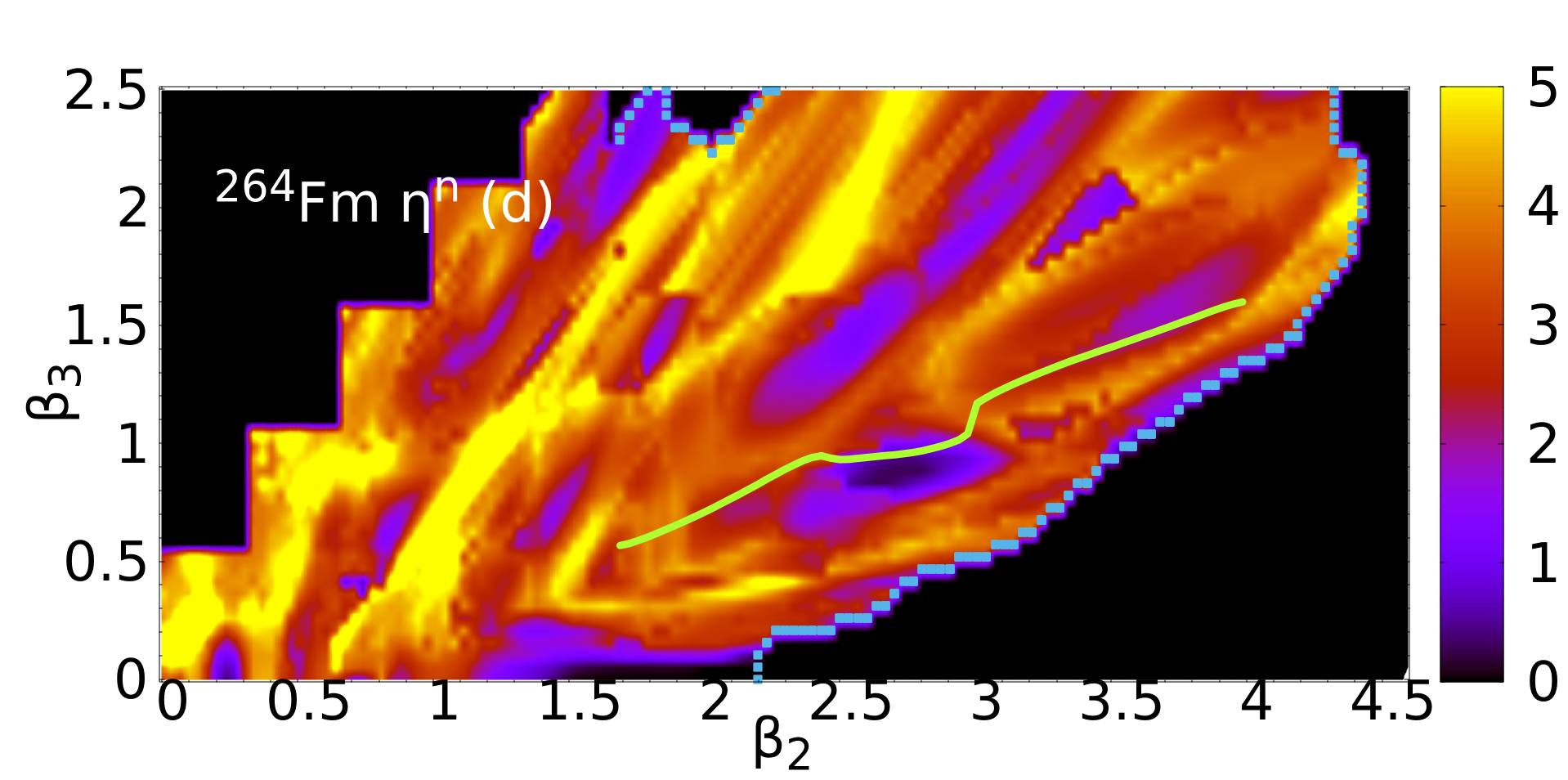} \\
   \includegraphics[width=0.49\linewidth ]{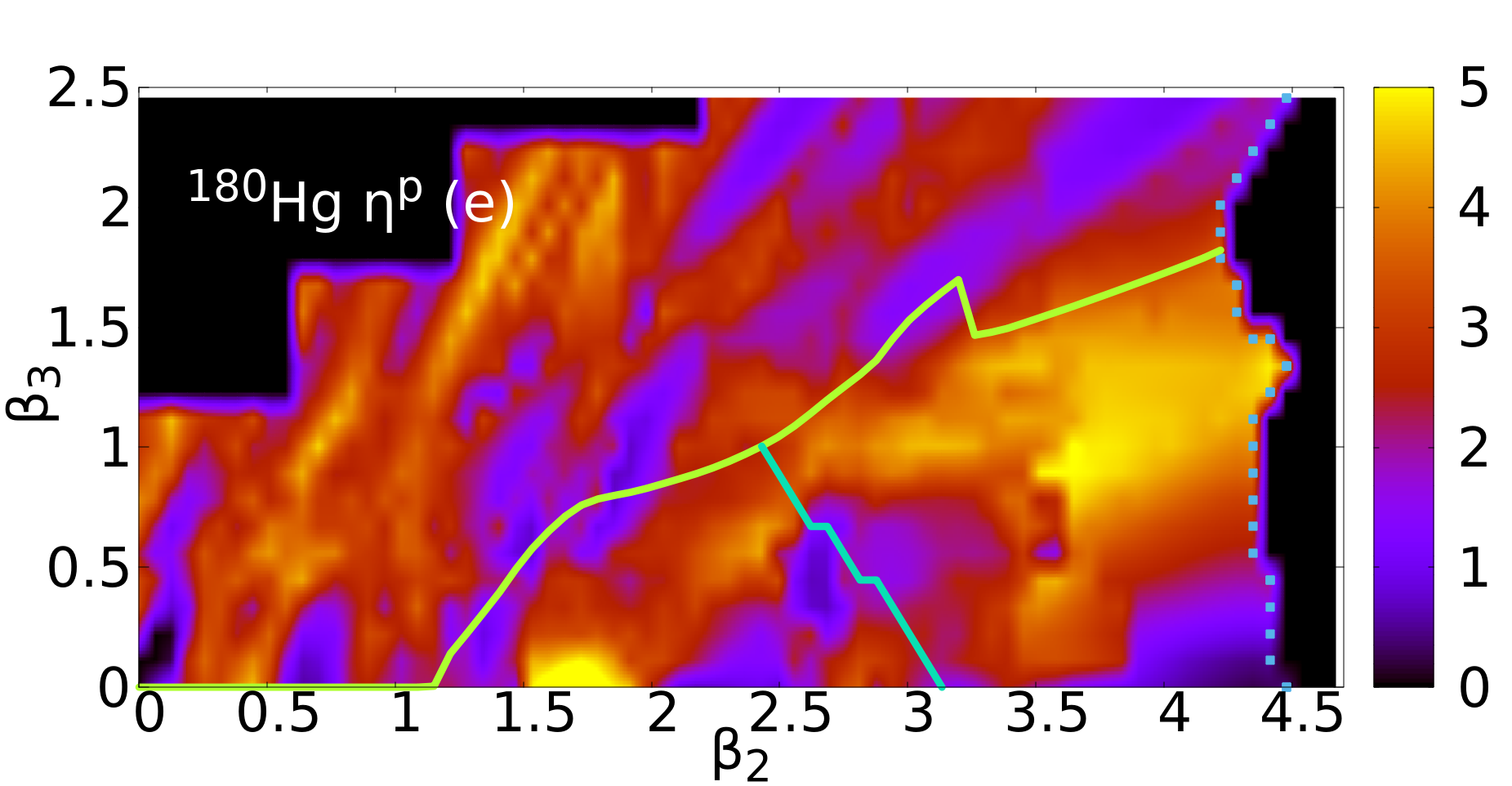} 
   \includegraphics[width=0.49\linewidth ]{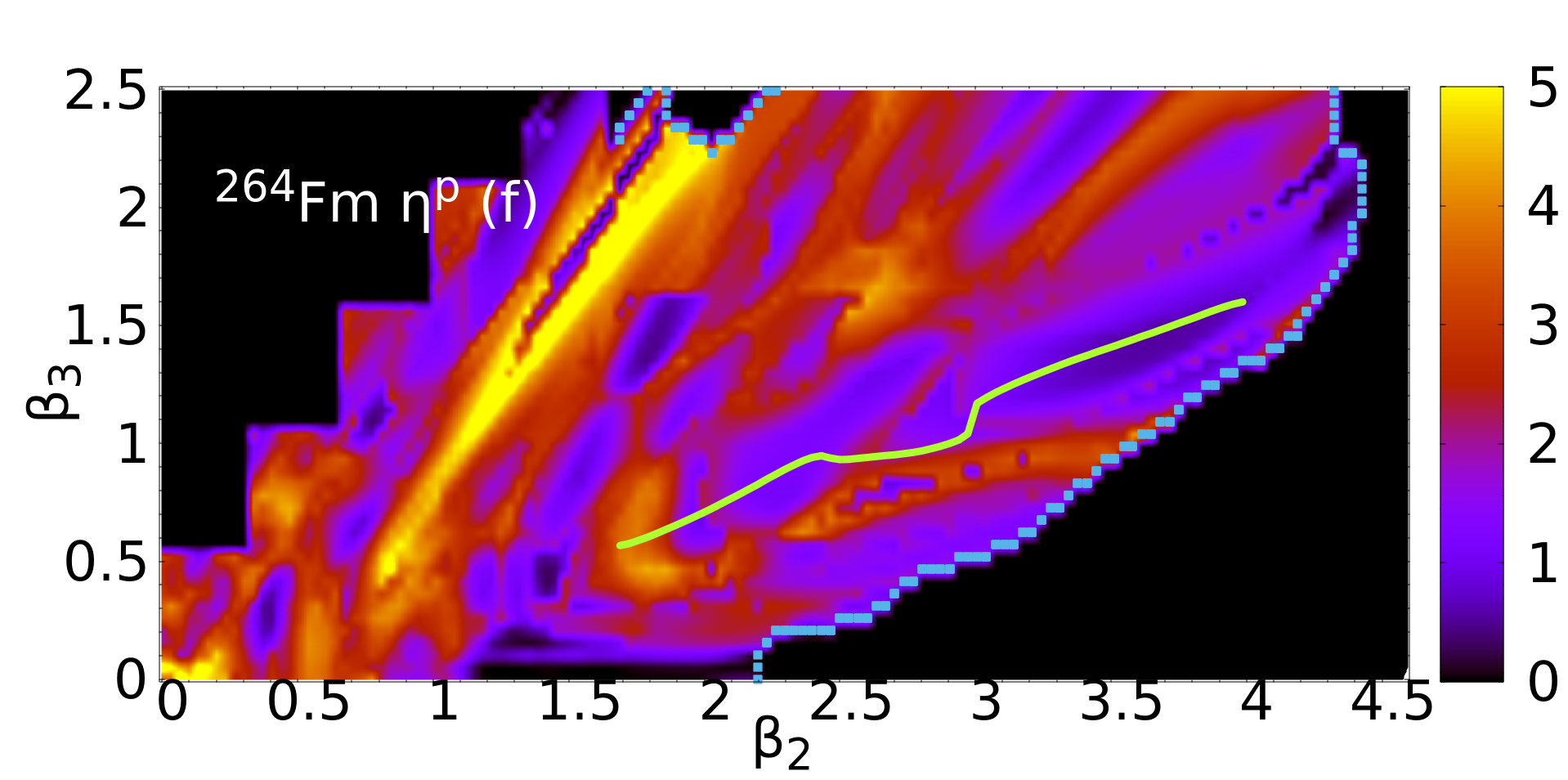} \\
\caption{\label{fig::PESgap} Panels (a) and (b): quadrupole-octupole moment constrained 
potential energy surfaces in MeV for $^{180}$Hg and $^{264}$Fm respectively. Spatial 
densities for the last configuration before scission of the 1D path are displayed in 
the inner panels. Panels (c) and (d): neutron smoothed level densities around the Fermi 
level. Panels (e) and (f): same as (c) and (d) for protons. For all the panels the 
asymmetric paths are depicted in yellow lines and the green line represents the 
transitional valley. The scission lines are given by the blue squares.
}
\end{figure*}

To begin the examination of the fission processes of $^{180}$Hg and $^{264}$Fm, we first analyze the PES generated from ($\beta_2$,$\beta_3$)-constrained HFB calculations. The findings are presented in panels (a) and (b) of Fig.~\ref{fig::PESgap}. A yellow line represents the path of minimum energy that links the ground state of the system to the scission point, as defined in Sec.~\ref{sec::framework}. It is observed that, at high $\beta_3$ asymmetry, the scission lines for both nuclei are positioned at significant elongations ($\beta_2 \ge 4$), indicating that each nucleus produces highly deformed fragments in their asymmetric fission modes. For $^{264}$Fm, early symmetric fission at $\beta_2 \sim 2.2$ occurs, cutting off the fission valley at a lower elongation compared to $^{180}$Hg. In the case of $^{264}$Fm, this symmetric fission leads to the formation of two doubly magic spherical $^{132}$Sn nuclei. 
While one might anticipate a similar outcome for $^{180}$Hg, yielding two spherical $^{90}$Zr fragments, symmetric scission in $^{180}$Hg occurs at a much larger elongation ($\beta_2 \sim 4.4$), resulting in two significantly deformed fragments.
\\

We now utilize the sld as defined in Eq.~\eqref{eq::eta} to explore how the PES are influenced by compound nucleus  and prefragments shell effects. This method offers a significant advantage in distinguishing between neutron and proton shell effects, as illustrated in panels (c) to (f) of Fig.~\ref{fig::PESgap}. Building on the discussion in Ref.~\cite{BernardEPJA23}, regions of low sld are distributed across the PES, with the 1D asymmetric paths influenced by a number of distinct low sld values for each isospin.
\\

\subsection{$^{180}$Hg sld}\label{sec::Hgsld}

For both nuclei, asymmetry is initiated at $\beta_2 \sim 1.2$, well before the prefragments or neck begin to form. 
In the $^{180}$Hg case, the onset of 1D path asymmetry occurs in a region of low sld for both protons and neutrons. Following the symmetric 
path would lead to regions of high 
proton sld at $\beta_2 \approx 1.6$, followed by high neutron sld at 
$\beta_2 \approx 2.5$. These areas are marked by level crossings near the 
Fermi level, contributing to an increase in mean field energy. In contrast, 
low sld along the nascent asymmetric valley tends towards more stable local 
energy conditions. Notably, around $\beta_2 \approx 2.5$, a new local valley appears, delineated by the green line in Fig.~\ref{fig::PESgap}, which 
returns from an asymmetric configuration to a symmetric configuration in a transitional valley. 
This valley is located in a zone of low proton sld, offering $^{180}$Hg the 
opportunity to reach  a symmetric fission configuration from the asymmetric path. The possibility of this outcome is investigated further in Section \ref{section::1Dpaths}.

A discontinuity in $\beta_{3}$ is noted at $\beta_2 \sim 3.25$ along the asymmetric path. Although methodologies exist to circumvent this issue (see, e.g. \cite{LauPRC22,lasseri2024,Carpentier2024}), we opt to proceed with the analysis post-discontinuity to scission, without compromising the overall findings of the study. In this particular section of the path, we pinpoint 
a stabilized pair of prefragments, $^{98}$Ru/$^{82}$Kr. This pair, 
 obtained without accounting for beyond mean field effects such as symmetry restoration, dynamic pairing, or dissipation (referenced in Ref.~\cite{bender2020,GiulianiPRC14,bernard2019,bernard2011,sierk2017, sadhukhan2016, randrup2011}), aligns closely with the central points of experimental fission yield peaks (Ref.~\cite{AndreyevPRL10,elseviers2013,NishioPLB2015}) and is in agreement with the Skyrme mean field approach presented in the supplemental material of Ref.~\cite{ScampsPRC19}.
\\
\begin{figure*}
  \centering
   \includegraphics[width=0.49\linewidth ]{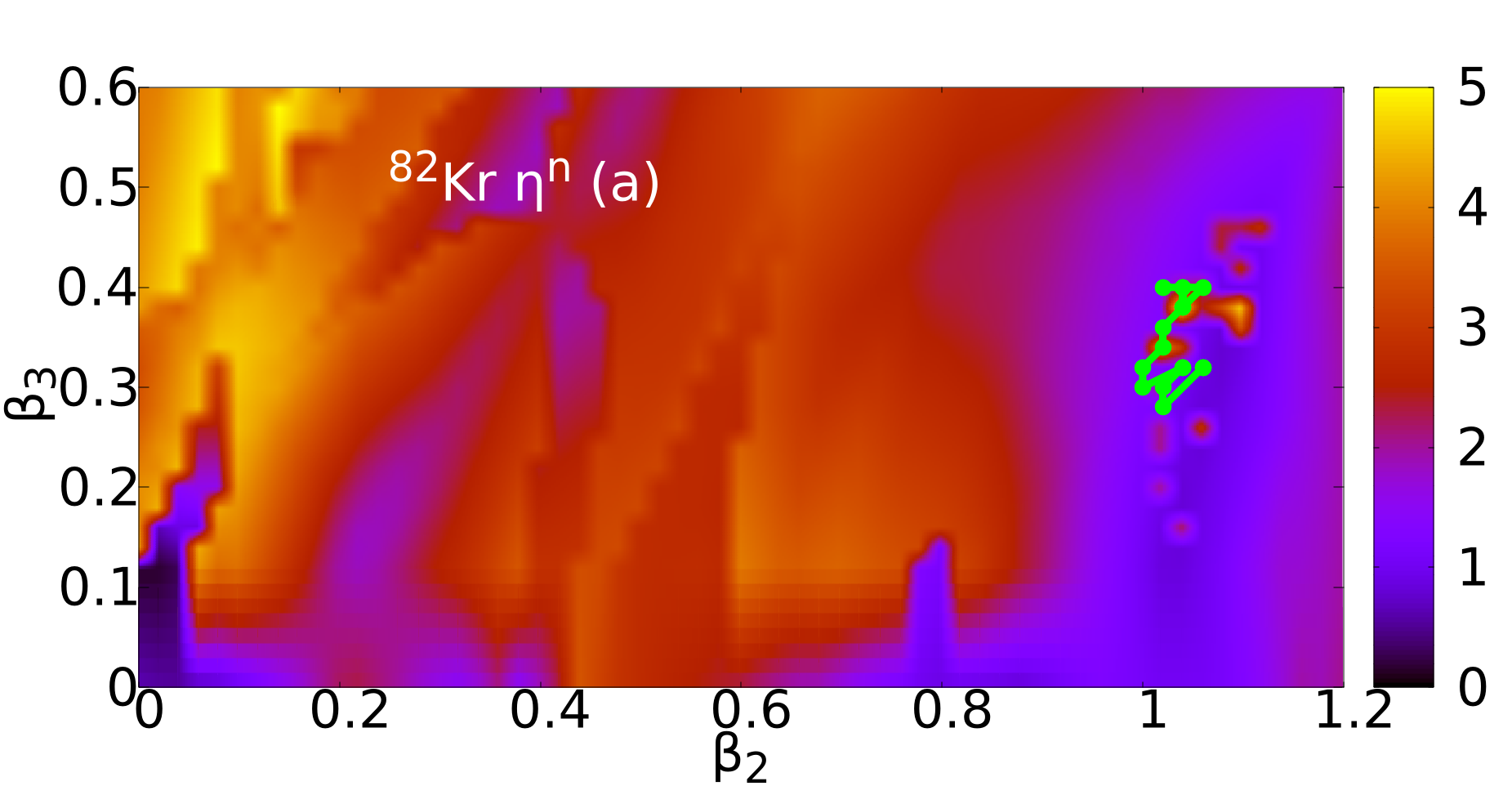}   
   \includegraphics[width=0.49\linewidth ]{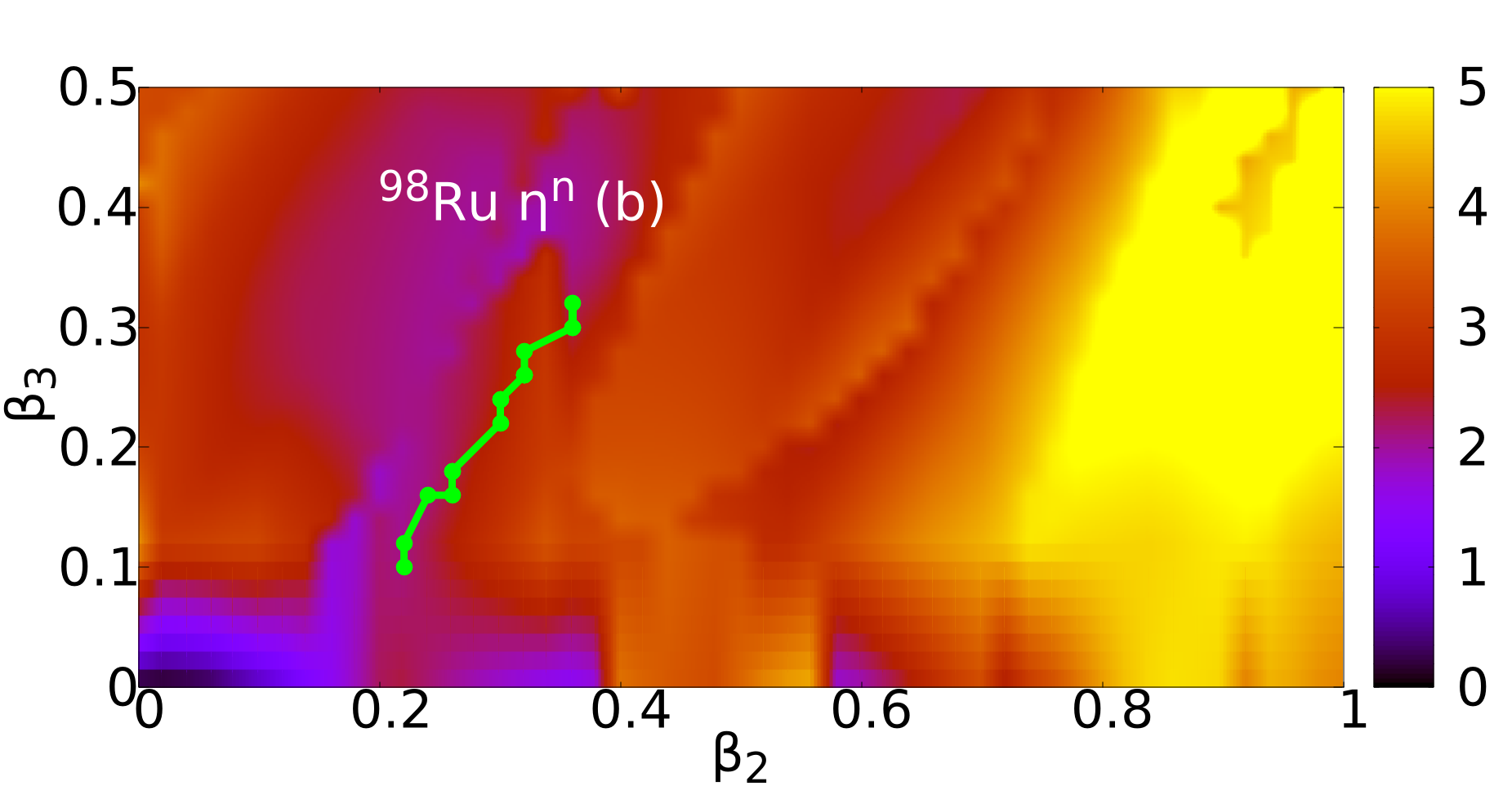} \\
   \includegraphics[width=0.49\linewidth ]{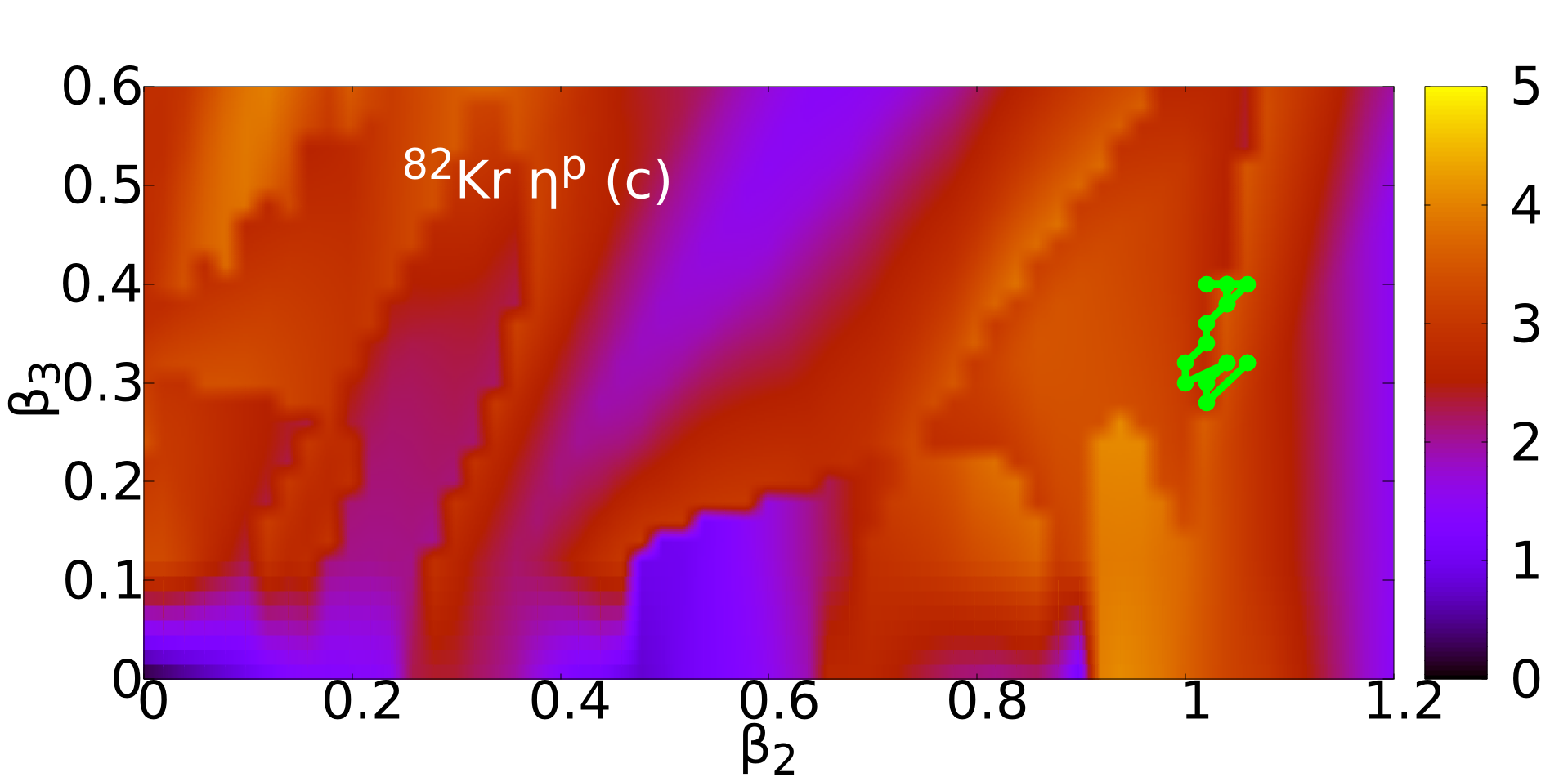}   
   \includegraphics[width=0.49\linewidth ]{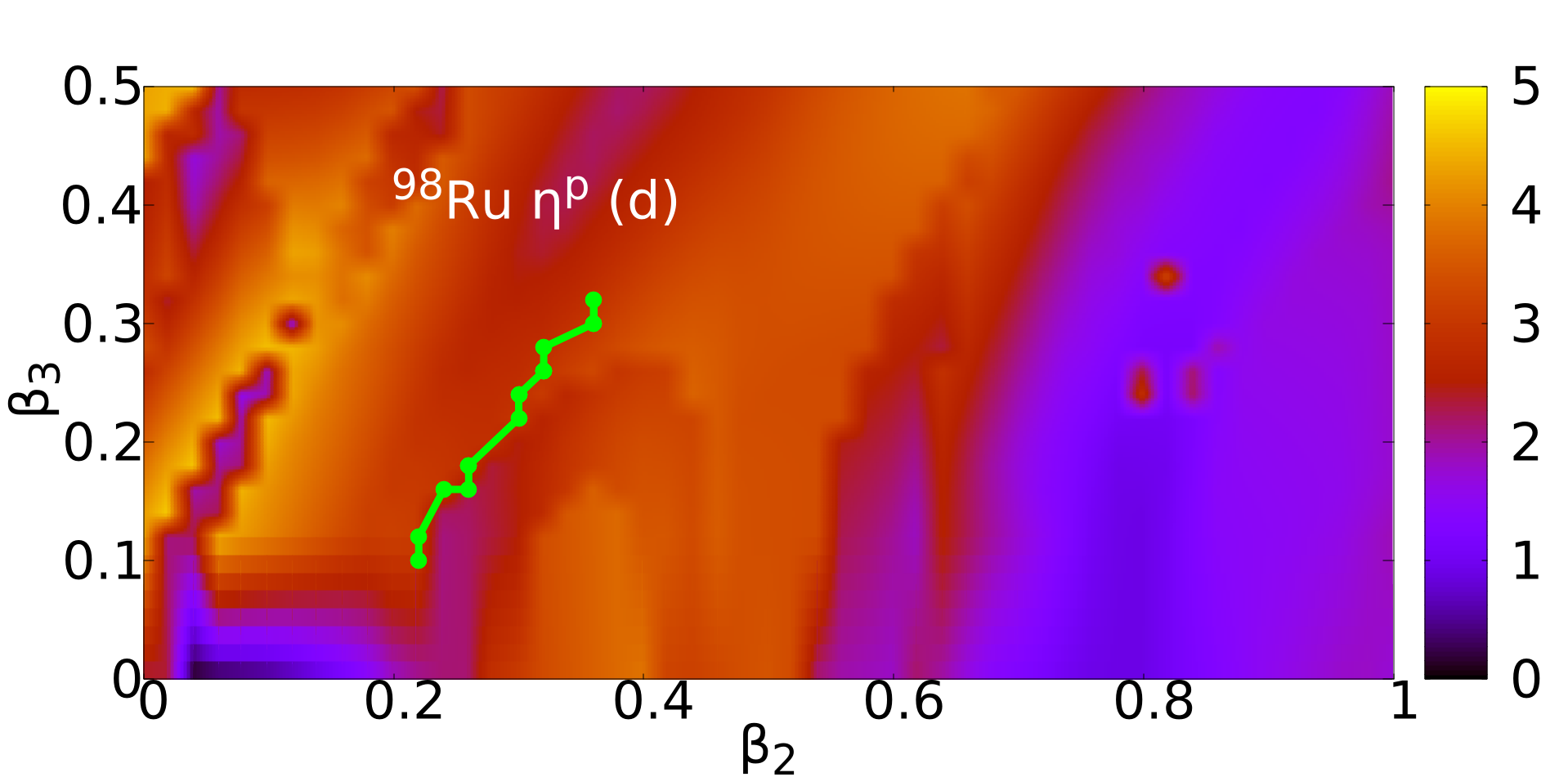} \\
\caption{Smoothed level densities of $^{82}$Kr and $^{98}$Ru. Panels (a) and (c) are the $^{82}$Kr neutron and proton sld respectively.
In panels (b) and (d) are plotted the neutron and proton sld for $^{98}$Ru respectively. The green lines with circle markers show the prefragment deformation during the final stages of the asymmetric $^{180}$Hg fission. The energy window around the Fermi level is fixed to 5~MeV for medium mass nuclei. Constraints are imposed on $Q_{20}$ and $Q_{30}$ and higher multipole moments are left unconstrained. 
}
\label{fig::sldfrag} 
\end{figure*}
We delve deeper into the analysis of the prefragment pair $^{98}$Ru/$^{82}$Kr. The methodology outlined in Sec.~\ref{sec::framework} not only facilitates the identification of prefragments but also delineates their respective deformations along the fission pathway of the compound nucleus in the prescission region. This enables the tracking of prefragment trajectories on their respective neutron and proton sld surfaces, as illustrated by green dotted lines in panels (a) to (d) of Fig.~\ref{fig::sldfrag}.

 The trajectories of prefragments on the neutron sld surface differ between the fragments; $^{98}$Ru begins with minor deformations and progressively increases both its quadrupole and octupole deformations. Conversely, $^{82}$Kr remains confined to a small area characterized by significant deformations. Additionally, $^{82}$Kr maintains its position within the low neutron sld region at a high quadrupole moment, whereas $^{98}$Ru does not display a clear presence in any low sld region. On the proton side, panels (c) and (d), the trajectories do not cross any noticeable low sld for both prefragments.
 From these observations, we deduce that the highly deformed shell effect at $N=46$, evident at the Fermi level of the lighter prefragment $^{82}$Kr, influences the later stage of $^{180}$Hg's asymmetric fission process.

It is noteworthy that this finding contrasts with the results presented in \cite{WardaPRC12b}, where the prefragments are predefined as a spherical $^{90}$Zr and a deformed $^{72}$Ge, maintaining the compound nucleus's Z/N ratio. Contrary to results in Ref.~\cite{WardaPRC12b} and our current sld analysis, the studies in Refs.~\cite{MollerPRC12,McdonnellPRC14} conclude that the shell correction energy, accounting for all shell effects of the single-particle spectrum, does not correlate with prefragment identities. 
\\
\begin{figure}
  \centering
   \includegraphics[width=0.7\linewidth ]{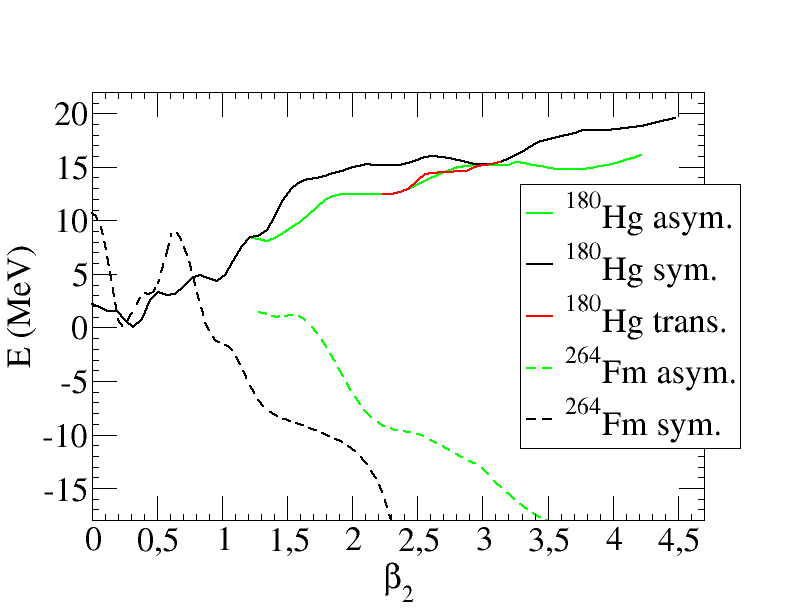} 
  \caption{Plots of the HFB energies along the 1D symmetric and asymmetric paths of $^{180}$Hg and $^{264}$Fm as functions of $\beta_2$.}
    \label{fig::1Dpaths}
\end{figure} 

\subsection{$^{264}$Fm sld}

The sld analysis applied to $^{264}$Fm supports the hypothesis that various proton and neutron shell effects significantly influence asymmetric fission. The findings are showcased in panels (b), (d), and (f) of Fig.~\ref{fig::PESgap}, corresponding to PES, neutron and proton sld surfaces, respectively. The path of asymmetric fission in $^{264}$Fm is predominantly guided by proton shell effects, with a notable deviation occurring in the range $\beta_2 \in [2.5,3.0]$ due to a neutron Fermi gap that momentarily alters the path from its expected trajectory before it returns to a region of low proton sld.

It is important to note that the 1D path does not extend to the scission line as the local PES becomes convex around $\beta_2 \sim 4$, causing the disappearance of the local valley. On the symmetric pathway, the emergence of strong low sld for both protons and neutrons around $\beta_2 \sim 1.2$ signifies the formation of doubly magic $^{132}$Sn prefragments, thereafter shaping a well-defined symmetric valley on the PES all the way to scission.

Interestingly, both the symmetric and asymmetric paths seem influenced by dual $^{132}$Sn prefragments. In the asymmetric scenario, the prefragments identified along the final proton low sld section (starting at $\beta_2 \sim 3.0$) comprise a spherical $^{132}$Sn and a deformed $^{132}$Sn counterpart. This is depicted in the density plot in the inner panel of Fig.~\ref{fig::PESgap}(b) for the last configuration before scission along the 1D path, where the position of the density minimum at the far left of the neck is likely to be  caused by the significant shell effects of the almost spherical left $^{132}$Sn fragment.

Exploration near the scission line reveals quadrupole-octupole deformed heavy nuclei around $Z\sim 54-56$ and their complementary pairs, such as $^{142}$Xe/$^{122}$Pd and $^{146}$Ba/$^{118}$Ru. These couples are located in the region of a proton shell effect around $(\beta_2,\beta_3)=(4.3,2.0)$ (see panel~(f) of Fig.~\ref{fig::PESgap}) after the end of the 1D path in the locally convex region of the PES.
As in the case of $^{258}$Fm \cite{hulet1986}, the associated deformed shell effects might induce a small shift in yields towards the asymmetric fission mode.

\subsection{PES/sld interplay}\label{section::1Dpaths}

For a deeper understanding of the interaction between sld and PES, the energies from HFB calculations for both symmetric and asymmetric paths are plotted against their elongation in Fig.~\ref{fig::1Dpaths}. It has been verified that introducing triaxiality does not reduce the barrier heights for the symmetric paths. 

The symmetric path for $^{264}$Fm has a noticeably lower energy compared to the asymmetric path. At significant elongations, both paths for $^{264}$Fm show some inflection points which, when close to scission, may be analyzed in terms of prefragment characteristics. An initial inflection at $\beta_2 \sim 1.2$ along the symmetric path marks the emergence of low sld for both protons and neutrons depicted in Fig.~\ref{fig::PESgap}, corresponding to the formation of spherical $^{132}$Sn prefragments. A subsequent inflection at scission, with $\beta_2 \sim 2.2$, leads to a change in the energy slope, caused by Coulomb repulsion between the emerging fragments.

In contrast, the asymmetric path for $^{264}$Fm is found to be several MeV higher in energy than the symmetric path, particularly as it contends with the formation of two spherical $^{132}$Sn prefragments that lower the symmetric valley. The presence of one nascent spherical $^{132}$Sn (the other one being elongated), responsible for the appearance of the low proton sld in Fig.~\ref{fig::PESgap}(f), contributes to a final inflection point at $\beta_2 \sim 3.0$.

In the case of $^{180}$Hg, both the symmetric and asymmetric fission paths increase in energy as they approach highly elongated scission configurations. A few MeV energy difference between these paths indicates a preference for an asymmetric scission outcome. The transitional valley that bridges the asymmetric and symmetric paths is also drawn in red in Fig.~\ref{fig::1Dpaths}. Notably, this connection occurs at a deformation where the symmetric and asymmetric paths are nearly equivalent in terms of mean field energy. The stability of this characteristic has been verified in the present work using the SLy4 Skyrme interaction within the \textsc{SkyAx} mean field code \cite{ReinhardCPC21}. Thus there is no significant extra energy cost between the symmetric and asymmetric configurations at $\beta_2 \sim 3.1$.

\section{$\beta_{2}-\beta_{4}$ study of PES and sld}
\label{sec::PESandgapS}

From the $\beta_{2}-\beta_{3}$ analysis in the previous section, we understand that $^{264}$Fm can achieve energetically favorable configurations incorporating $^{132}$Sn nuclei along both symmetric (with two spherical fragments) and asymmetric paths (with one spherical and one deformed fragment). This prompts the question of whether $^{90}$Zr might exhibit similar characteristics within $^{180}$Hg's fission dynamics. Given the negligible energy difference in returning to its symmetric valley, it is intriguing that $^{180}$Hg does not undergo symmetric scission around $\beta_2 \sim 3.1$, potentially yielding either slightly deformed $^{90}$Zr nuclei or a pair consisting of a spherical $^{90}$Zr and its deformed complement.

Given that the asymmetric scission point is at a higher energy, the presence of the transitional valley might offer $^{180}$Hg an opportunity for a symmetric contribution to the mass and charge yields. With this possibility in mind, we will delve deeper into the characteristics of the symmetric valley.
\\
\begin{figure*}
  \centering
   \includegraphics[width=0.49\linewidth ]{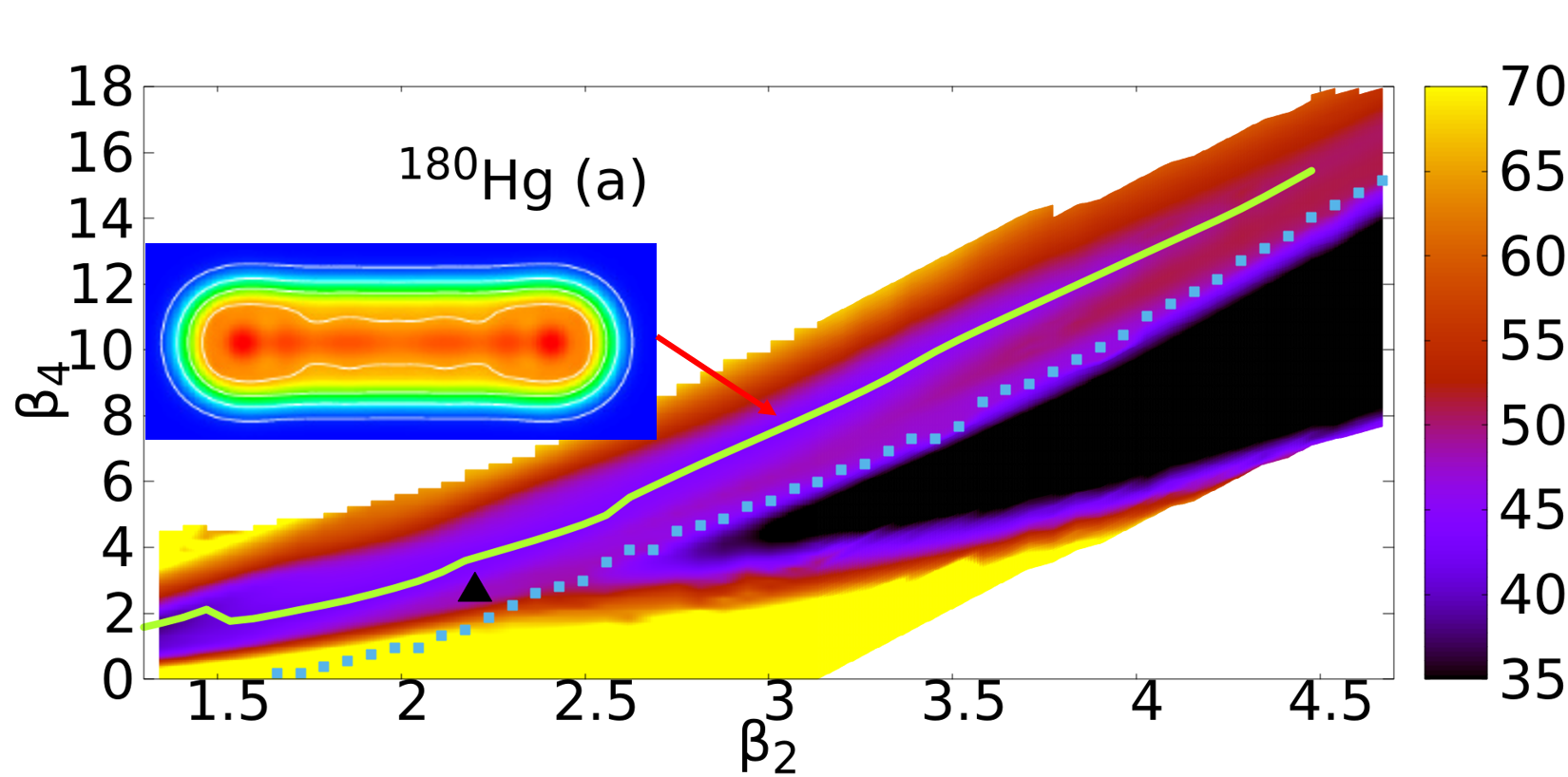}   
   \includegraphics[width=0.49\linewidth ]{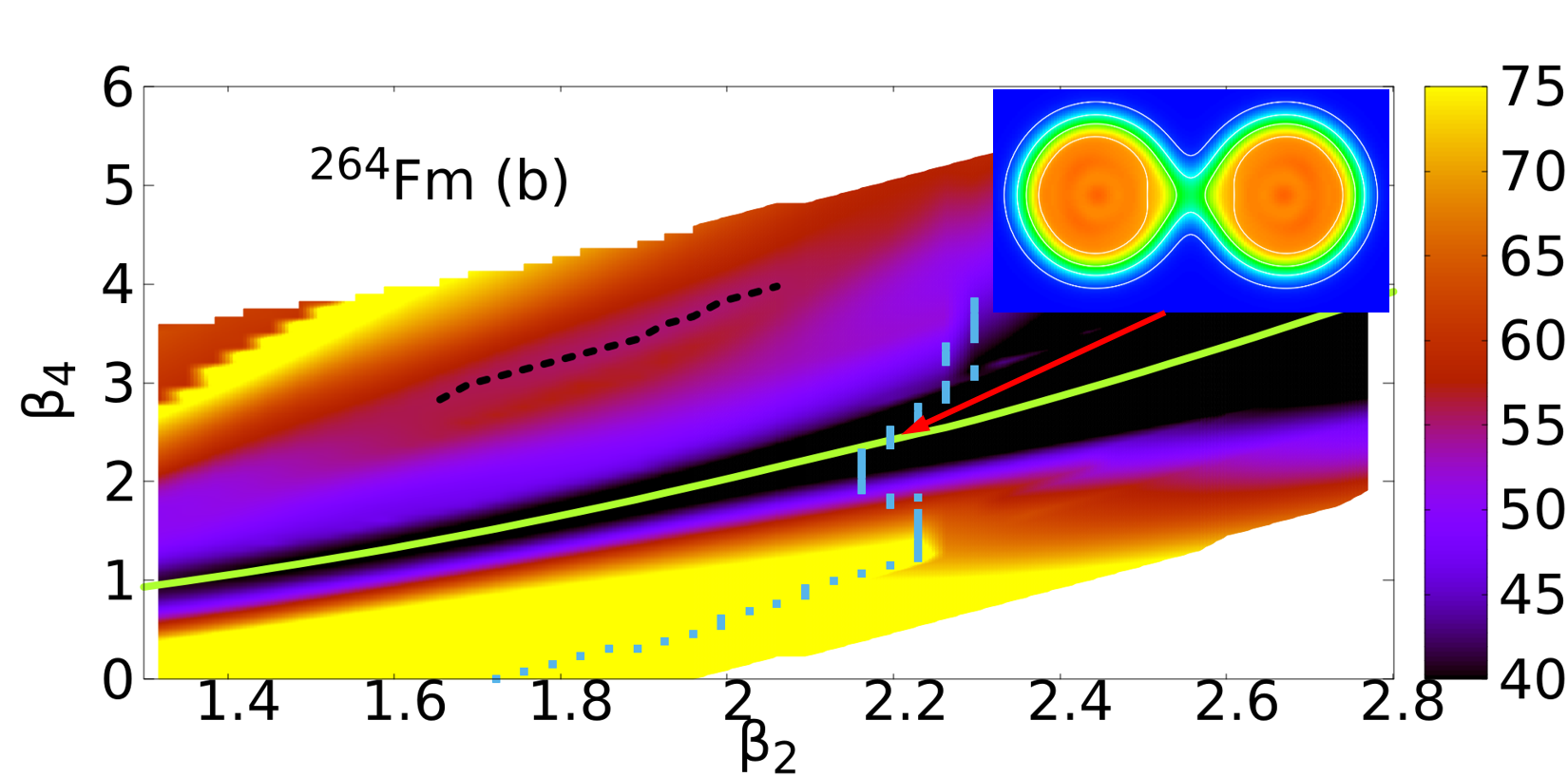} \\
   \includegraphics[width=0.49\linewidth ]{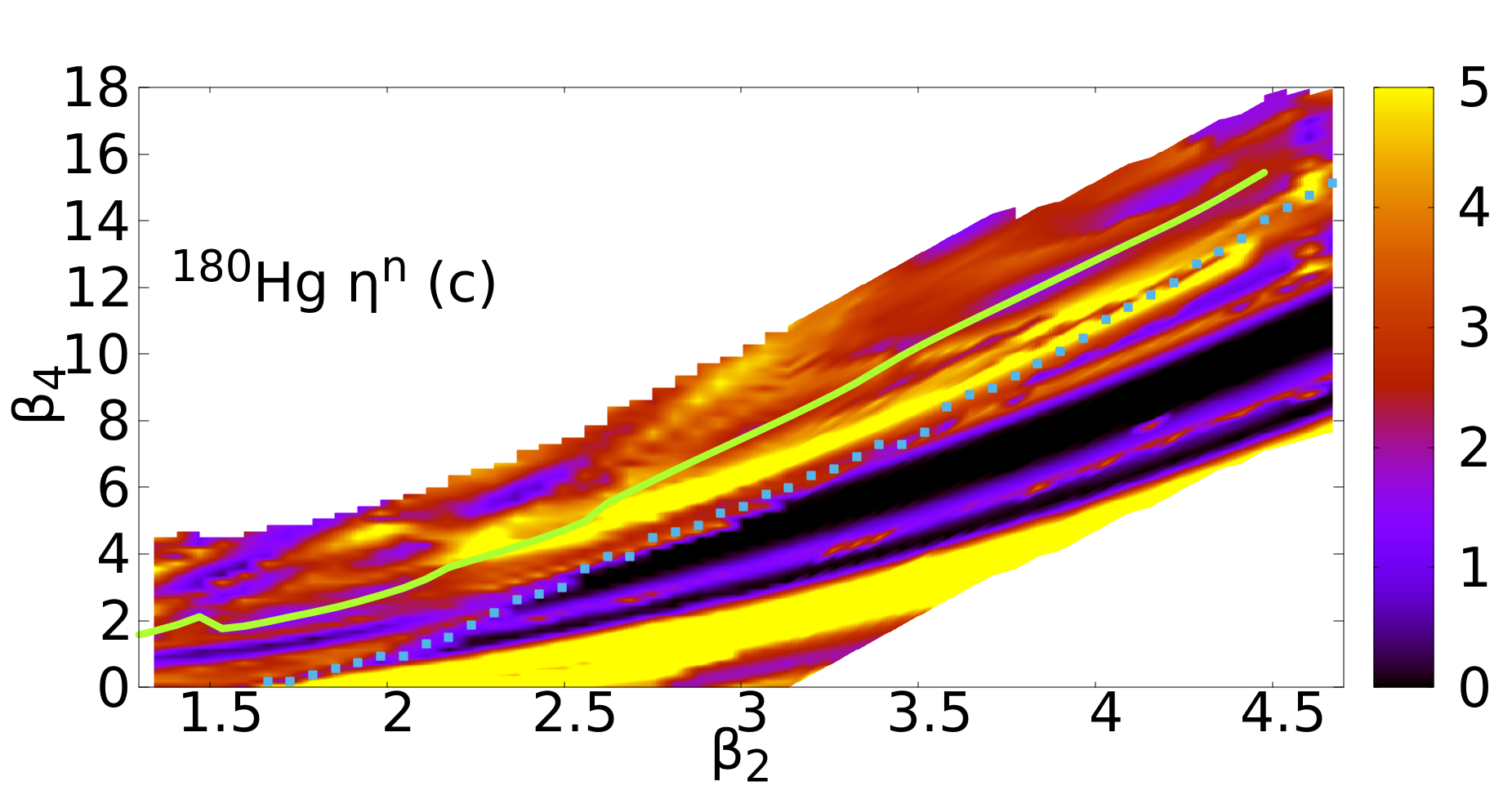}   
   \includegraphics[width=0.49\linewidth ]{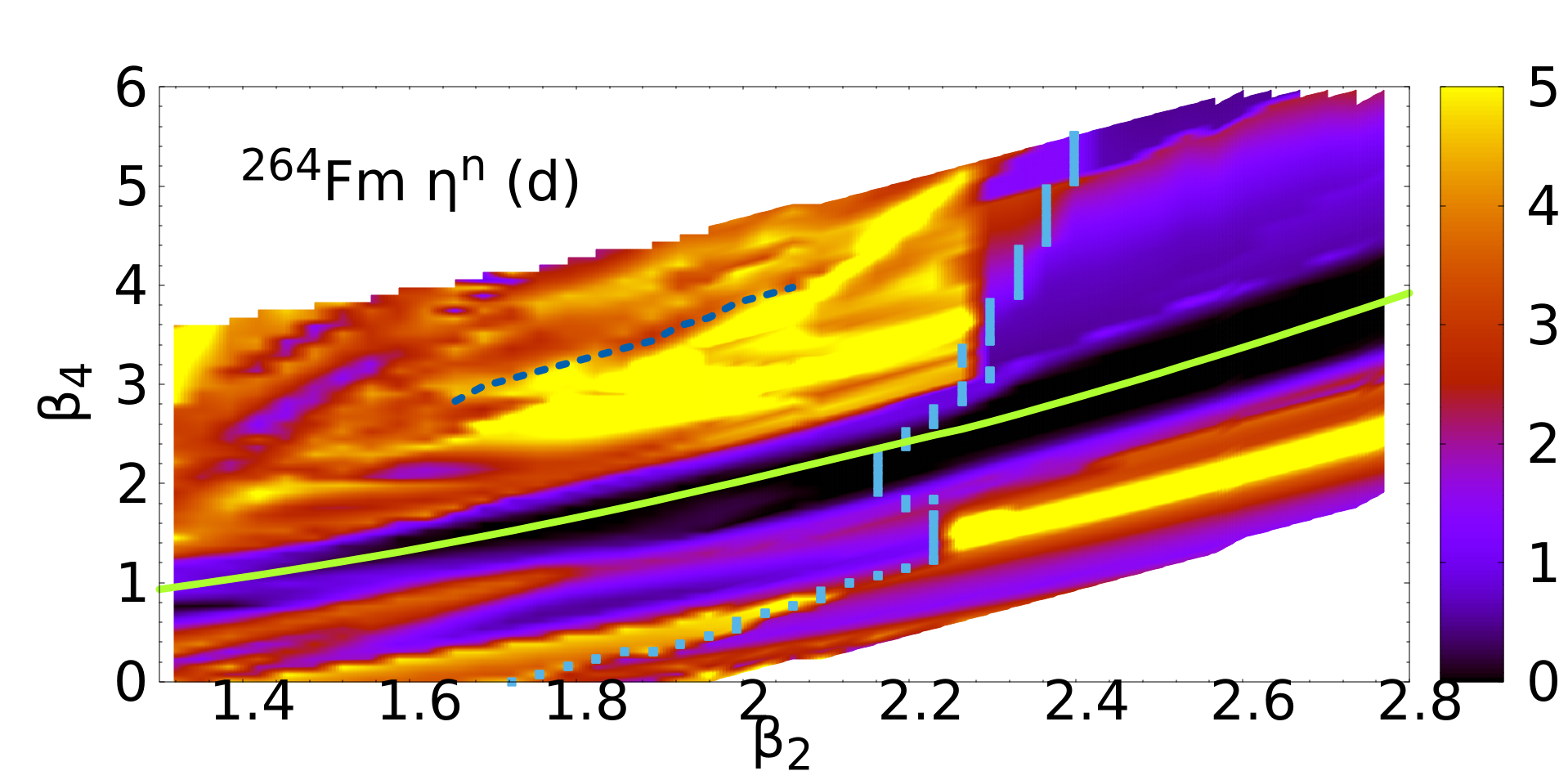} \\
   \includegraphics[width=0.49\linewidth ]{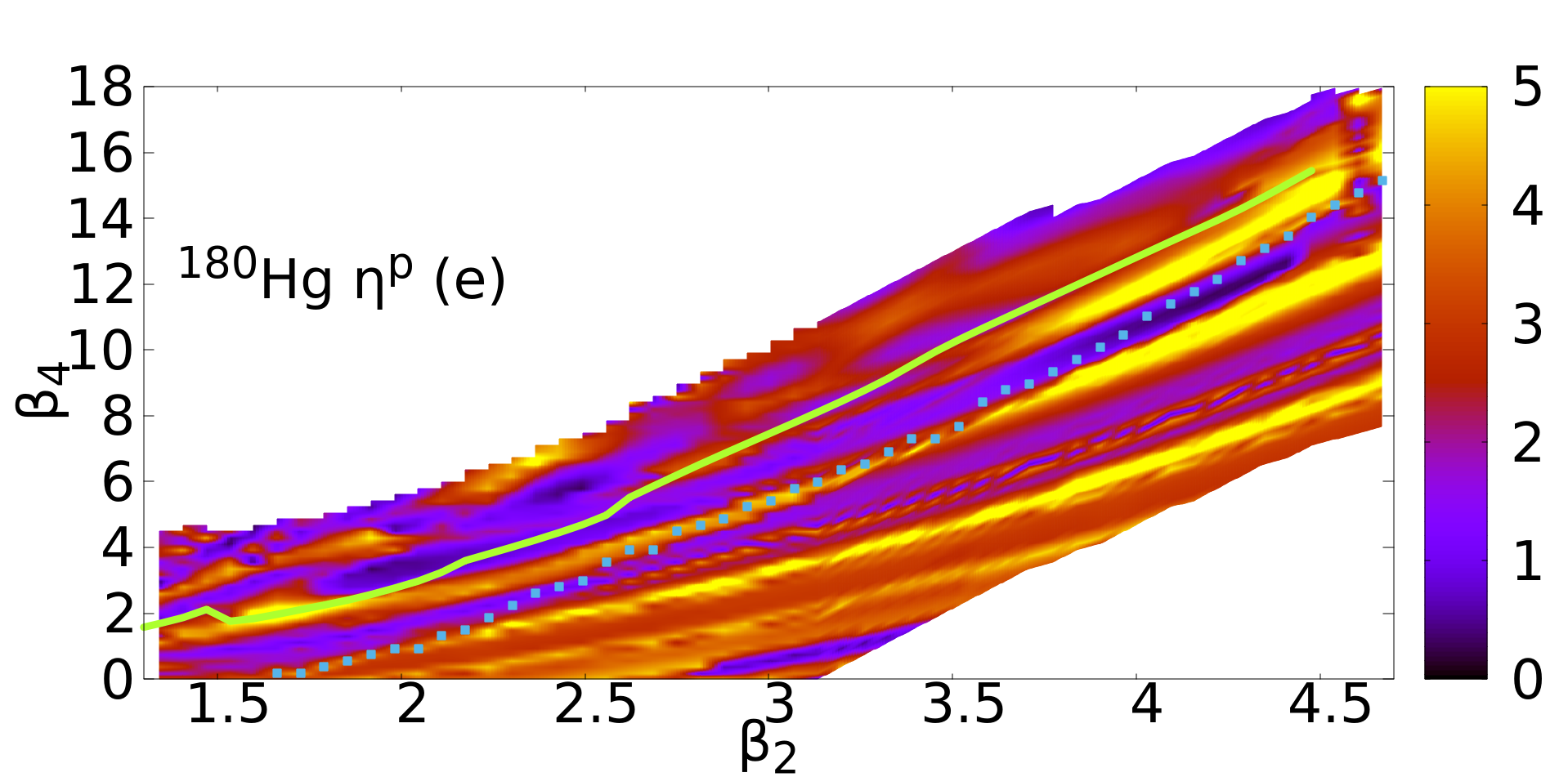}   
   \includegraphics[width=0.49\linewidth ]{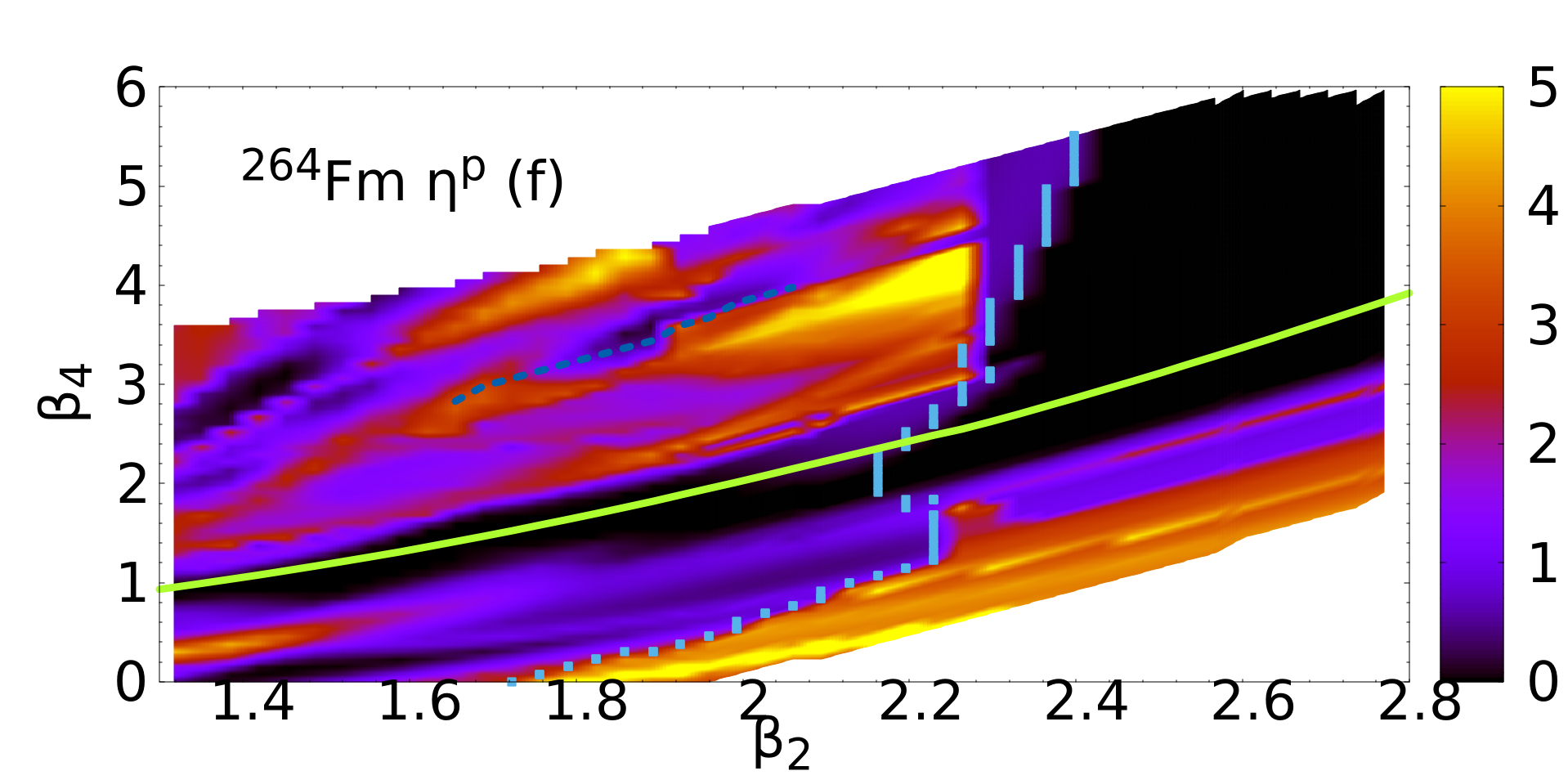} \\
\caption{Quadrupole-hexadecapole moment constrained potential energy 
surfaces for $^{180}$Hg (a) and $^{264}$Fm (b) with colours depicting energies in MeV. The octupole moment is set to zero.
Smoothed level densities around the Fermi level are shown for neutrons in panels (c) and (d), and for protons in panels (e) and (f). 
In all plots the symmetric paths are drawn with  green lines, and the scission lines with dotted blue lines. The dashed line in the $^{264}$Fm panels represents the secondary valley. The triangle in (a) is placed at the same $(\beta_2,\beta_4)$ position as the $^{264}$Fm scission configuration shown in the inset of (b). }
\label{fig::PESsym}
\end{figure*}
In Fig.~\ref{fig::PESsym} the PES and sld plots for $^{180}$Hg and $^{264}$Fm are presented for symmetric fission, beginning at well deformed configurations ($\beta_2 = 1.3$) and going beyond scission. Quantities are plotted against quadrupole and hexadecapole moments with a fixed zero octupole moment ($\beta_3=0$). 
Solid lines are the symmetric 1D fission, corresponding to the 
$\beta_{3}=0$ lines of Fig.~\ref{fig::PESgap} PES, while blue dots define the scission lines.  Results are discussed in the next subsections.

\subsection{$^{180}$Hg $\beta_{2}-\beta_{4}$ PES}

In the $^{180}$Hg case, the symmetric path extends towards higher $\beta_4$ values, indicating a progressively elongating nucleus while consistently staying in the bottom of the valley until reaching approximately $\beta_{2} \sim 4.5$. The subsequent point along this trajectory (not depicted) would transition into the fusion valley. As shown in the inner panel (a) of Fig.~\ref{fig::PESsym}, the overall spatial density distribution reveals that the compound nucleus at $\beta_2 \sim 3.1$ still lacks the characteristic neck formation typically associated with a dinuclear system near scission.

An energy ridge separating the fission path and the scission line persists along the entire PES, effectively creating a fission barrier of a few MeV that diminishes with increasing elongation. Consequently, this ridge prevents $^{180}$Hg from undergoing scission in a compact mode. This ridge is present at the exit point of the transitional valley depicted in Fig.~\ref{fig::PESgap}, which reaches the 1D symmetric path around $\beta_2 \sim 3.1$. To illustrate this point further, a mean field energy slice at $\beta_2 = 3.1$ is provided in the Appendix in Fig.~\ref{fig::slice100b}, highlighting an approximate $4~$MeV barrier separating the symmetric path from the fusion valley.

\subsection{$^{264}$Fm $\beta_{2}-\beta_{4}$ PES}

In panel (b) of Fig.~\ref{fig::PESsym}, the symmetric path of $^{264}$Fm consistently maintains low $\beta_4$ values, reaching the scission line at low elongation in a highly compact configuration. It is noteworthy that the 1D path goes through the scission line without any noticeable energy drop. 
At scission, two spherical $^{132}$Sn fragments are found
as depicted in the inner panel. 
Along the symmetric path, the compound nucleus follows the 
minima of the single-particle level density (sld) for both neutrons and protons.  They are displayed in Fig.~\ref{fig::PESgap}(d) for neutrons and Fig.~\ref{fig::PESgap}(f) for protons, respectively. Prefragments emerge well before reaching scission, remaining spherical from their 
inception until the moment of scission. Additionally, it is worth mentioning the presence of a secondary symmetric valley at higher $\beta_4$ values and higher energy 
(approximately 18 MeV) represented as the dashed line in panels (b), (d) and (f). This secondary path for $^{264}$Fm corresponds to the primary symmetric path observed for $^{180}$Hg in Fig.~\ref{fig::PESsym}(a) in terms of ($\beta_2,\beta_4$) deformations, albeit disappearing 
at smaller elongations. 
As observed in $^{180}$Hg, the large $\beta_4$ values in 
$^{264}$Fm prevent the introduction of any prefragments at this stage due to the 
undefined neck structure. 

\subsection{$^{264}$Fm $\beta_{2}-\beta_{4}$ sld}

Panels (d) and (f) in Fig.~\ref{fig::PESsym} display the neutron and proton sld for $^{264}$Fm respectively, both before and after the scission line.
As already shown in Fig.~\ref{fig::PESgap}(d) and Fig.~\ref{fig::PESgap}(f), the 1D symmetric path closely follows the distinct low neutron and proton sld, indicative of the emergence of $^{132}$Sn prefragments as discussed in Section~\ref{sec::PESandgapA}. Notably, the secondary 1D path at higher $\beta_4$ is situated within a region of low proton sld, which disappears around $\beta_2\simeq2$ and $\beta_4\simeq4$.

\subsection{$^{180}$Hg $\beta_{2}-\beta_{4}$ sld}
For $^{180}$Hg, the scenario regarding sld is notably different, as depicted in panels (c) and (e) of Fig.~\ref{fig::PESsym} for neutrons and protons respectively. In contrast to the case of $^{264}$Fm, the symmetric path of $^{180}$Hg does not closely align with the minima of neutron or proton low sld at low $\beta_4$, reflecting the fact that the path does not follow a well defined minimum energy valley in the $\beta_2-\beta_3$ PES.

Along the energy ridge that separates the 1D path from the scission line, a stretched proton low sld and a neutron high sld coincide. The elevated neutron sld arises from several single-particle intruder states, resulting in a sudden restructuring of the $^{180}$Hg nucleus's configuration between the scission line and the 1D path. As the states on this ridge approach the scission line, they may be regarded as forming a prefragments-plus-neck structure.

The transition from the fission valley to the fusion valley is depicted in terms of single-particle energies from the compound nucleus and the symmetric fragment in Fig.~\ref{fig::SPE_Hg_Zr} of the Appendix. It can be seen that neutron and proton intruder states in the ridge region correlate with those in the deformed $^{90}$Zr fragment. The abrupt structural shift at the saddle point marks the transition from a dinuclear-plus-neck structure to a stretched compound nucleus configuration. The latter state cannot be interpreted in terms of prefragments.

\subsection{Valley layering}

In this study, PES are constructed under the adiabatic approximation, where each point on the PES represents the mean field energy minimum of the compound nucleus under the considered deformation constraints.
For most of the heavy nuclei experiencing fission there is a competition between the fission and fusion valleys, the fusion valley being located at lower energies at large deformations.
A strict adiabatic approximation would make the compound nucleus scission where the two valleys cross each other in a compact mode. However constraining the $\beta_{4}$ operator as done in Fig.~\ref{fig::PESsym} or Fig.~\ref{fig::slice100b} highlights an energy barrier between the two valleys for $^{180}$Hg preventing the system to scission. Instead the calculations are extended in the fission valley for larger $\beta_2$ deformations until the barrier height becomes negligible. Thus for $^{180}$Hg, the scission line is associated with  a transition between these two types of valleys, involving a drastic change in nucleus structure in this region along with a big energy drop. Conversely, no such abrupt change is observed for $^{264}$Fm in the region where the symmetric 1D path intersects the scission line.
Using the overlaps between two mean field states as a measure of their similarity, we determined that the overlaps along the 1D path between two consecutive HFB states remain high, even as the path crosses the scission line (greater than 90\%). This suggests a very uncommon feature that the segment of the symmetric path leading up to the scission line is continuous with the fusion valley. Then one could see the $^{264}$Fm prescission area as an extension of the fusion valley.

 One might ask in which extend these different behaviors are connected to the respective symmetric fragments. To further investigate this, we examine, for each 
compound nucleus, two distinct states with identical ($\beta_2, \beta_4$)
deformations. One is a fission like configuration (strong neck) and is denoted as $C_{\text{fis}}$, while the other is a fusion-like configuration (smaller neck) and is labeled as 
$C_{\text{fus}}$. We select the deformation corresponding to the $^{264}$Fm scission 
point along its symmetric path as a reference, for which the spatial density is plotted in the inset in Fig.~\ref{fig::PESsym}(b). This defines $C_{\text{fus}}$($^{264}$Fm). Then, we employ the same ($\beta_2, \beta_3=0, \beta_4)$ deformation to define the $C_{\text{fis}}$ configuration for $^{180}$Hg, marked by the black triangle 
in Fig.~\ref{fig::PESsym}(a), positioned in the fission valley near the scission line. 
Then the fission configuration for $^{264}$Fm 
is achieved by forcing $^{264}$Fm to have the same shape than $^{180}$Hg in its $C_{\text{fis}}$ configuration. Doing so demands additional constraints on the nuclei's shapes: multipole moments $\beta_l$ are used up to $l = 6$, as well as the neck operator $Q_N$, in order to obtain the desired shape. The same process is applied to get $C_{\text{fus}}$($^{180}$Hg) from $C_{\text{fus}}$($^{264}$Fm). The $^{180}$Hg $C_{\text{fus}}$ HFB energy is found to be higher than the $^{180}$Hg $C_{\text{fis}}$ one, making the fission valley below the fusion one. Conversely the $C_{\text{fus}}$ energy is lower than $C_{\text{fis}}$ one for $^{264}$Fm making the layering between configurations inverted.

For each of the four configurations, we determine the prefragment deformations by identifying the lowest density in the neck as the separating point between prefragments. Subsequently, prefragment energies $E_{\text{frag}}$ and the Coulomb repulsion energy $E_{\text{CR}}$ between them are calculated using deformation constraints spanning from $\beta_2$ to $\beta_6$ for each prefragment. The energy distribution of the compound nuclei is then evaluated using the following decomposition:
\begin{equation}
E_{\text{HFB}} = 2E_{\text{frag}} + E_{\text{CR}} + E_{\text{diff}},
\end{equation}
which defines the difference energy $E_{\text{diff}}$. 
$E_{\text{diff}}$ would vanish for well separated fragments.
Table~\ref{tab::compaconf} provides details regarding the energy differences from the $C_{\text{fis}}$ to $C_{\text{fus}}$ configurations. 
\begin{table}[h]
\caption{\label{tab::compaconf}
Energy differences in MeV between the fission-like ($C_{\text{fis}}$) and fusion-like ($C_{\text{fus}}$) configurations for $^{180}$Hg and $^{264}$Fm.}
\begin{tabular}{@{}lllll@{}}
\toprule
Nucleus & $\Delta E_{\text{HFB}}$ & $\Delta E_{\text{CR}}$&
$2\Delta E_{\text{frag}}$& $\Delta E_{\text{diff}}$   \\
\midrule
$^{180}$Hg & -4.7 & +2.6 & +10.0 & -17.3 \\
$^{264}$Fm & +5.6 & +3.2 & +16.7 & -14.3 \\ 
\botrule
\end{tabular}
\end{table}

As the valley layering is inverted between $^{180}$Hg and $^{264}$Fm, their respective HFB energy differences exhibit opposite signs. The Coulomb repulsion differences are found to be small and are close to each other. These differences are found to be positive because the centers-of-mass of the prefragments are closer to each other in the $C_{\text{fus}}$ configuration. Conversely, the fragment and difference contributions vary significantly between the two systems. For $^{264}$Fm, the fragment contribution is relatively higher and, when combined with the Coulomb contribution, it dominates the difference energy, resulting in the fusion path lying below the fission one. In contrast, for $^{180}$Hg, the fragment energy is dominated by the difference energy, rendering the fission configuration the most bound state. Both energy differences are negative, indicating a gain in energy during neck formation. 

\begin{figure}
  \centering
   \includegraphics[width=0.7\linewidth ]{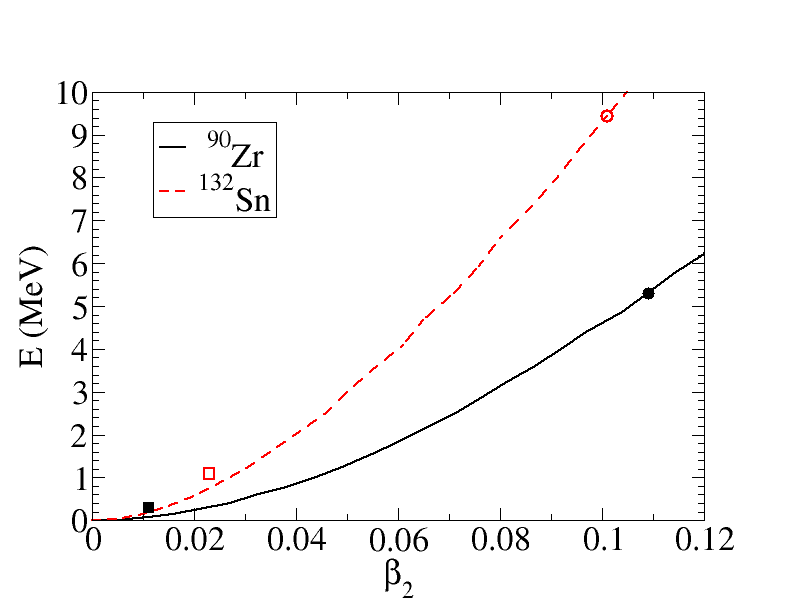}
\caption{\label{fig::PESfrag} Potential energy curves for $^{90}$Zr and $^{132}$Sn
at small deformations. See text for explanations.
The $C_{\text{fus}}$ ($C_{\text{fis}}$) configurations are represented by full black and open red squares (circles)
for $^{90}$Zr and $^{132}$Sn, respectively.}
\end{figure}

The potential energy curves of $^{132}$Sn and $^{90}$Zr at low $\beta_2$ deformation are presented in Fig.~\ref{fig::PESfrag}. Prefragment deformations for the $C_{\text{fus}}$ and $C_{\text{fis}}$ configurations are denoted by squares and circles, respectively. These potential energy curves are obtained by linearly relaxing the system from the $C_{\text{fis}}$ configuration to the ground state of the fragment. It is observed that $^{90}$Zr is relatively softer than $^{132}$Sn. Fragment deformations in the $C_{\text{fis}}$ configurations are similar, but the energy required to form $^{132}$Sn is significantly higher than that needed for $^{90}$Zr. This softer behavior of $^{90}$Zr reflects the $6.7$~MeV energy difference in Table~\ref{tab::compaconf} between the fragment contributions $2\Delta E_{\text{frag}}$, which is the main factor contributing to the change between both nuclei.

The question of whether the relative softness of $^{90}$Zr compared to $^{132}$Sn is sufficient to explain the inverted fission-fusion valley layering observed between $^{180}$Hg and $^{264}$Fm can be answered by supposing that  $^{90}$Zr has the same rigidity as $^{132}$Sn. Doing so reveals that $2\Delta E_{\text{frag}}$ increases by approximately $6.7$ MeV, which would be adequate to change the sign of $\Delta E_{\text{HFB}}$ ($\Delta E_{\text{HFB}}\simeq 2$ MeV). This argument holds under the assumption that the difference energy does not decrease significantly during the minimization of the overall mean field energy of the compound nuclei.

Consequently, it can be inferred that in the region of the symmetric valley, $^{90}$Zr is not rigid enough to enable a symmetric compact mode for $^{180}$Hg, regardless of the barrier heights at lower elongations. At large deformations this softness allows $^{180}$Hg to adopt configurations bound by a strong neck.

\section{Summary and conclusions}
\label{sec::conclu}
The low energy fission of $^{180}$Hg has been studied with the 
Hartree-Fock-Bogoliubov approximation and compared to the fission of $^{264}$Fm.
The smoothed level densities around the proton and neutron 
Fermi levels are analysed in the $\beta_2-\beta_3$ and 
$\beta_2-\beta_4$ planes, together with the corresponding potential energy surfaces.
Symmetric or asymmetric one-dimensional paths following energy valleys are
strongly correlated to shell effects associated with low sld.
For $^{180}$Hg, several proton and neutron low sld arise at small 
elongations and are responsible for the initial asymmetry of the 
1D path. These first shell effects are located at deformations too small to be correlated to any prefragment.
The 1D asymmetric path follows regions of low  neutron and proton sld identified as various 
shell effects as drivers of the most probable fission path.
At large deformations, the 1D asymmetric path encounters a discontinuity induced by a neutron shell effect that remains until scission. This shell effect is related to a strong $N=46$ neutron shell effect in the highly deformed light $^{82}$Kr prefragment.  
Additionally, a region of low sld gives rise to a transitional valley connecting
the 1D asymmetric path to the symmetric one without any extra cost in energy.

$^{180}$Hg is offered three opportunities to scission symmetrically but fails to provide any low energy significant symmetric mode.
Firstly, the compact mode is avoided due to a  symmetric fission barrier that is much higher in energy than its asymmetric counterpart. At larger elongations the
fusion and fission valleys are separated by an energy barrier, in contrast 
to the $^{264}$Fm scenario in which the fusion valley
remains lower in energy compared to the fission valley.
This difference is due to the relative softness of $^{90}$Zr
around sphericity compared to $^{132}$Sn.
An intermediate symmetric mode could have been possible by returning from asymmetry through the transitional valley. However, this path does not lead to a viable fission mode as the system reaches a 1D
symmetric path with high $\beta_4$ where $^{180}$Hg has a high elongation without prefragment or neck structure. 
This 1D path and the fusion valley are
separated by a energy ridge of a few MeV.
The stability of this low-energy stretched $^{180}$Hg state
prevails over the $^{90}$Zr fragment properties.
This 1D path finally rises in energy as the elongation increases,
making the elongated mode significantly higher in energy than the asymmetric
mode. 

During the descent from the saddle point to scission for both $^{180}$Hg and $^{264}$Fm, several shell
effects participate in guiding the wave function through
the potential energy surfaces. In particular the paths close to 
scission are influenced by shell effects of the prefragments.
The natures of the pairs of prefragments  
that come into play at very large deformations and the competition between them
should be analysed in more detail.
Whether several shell effects from several couples play sequential
roles on the way to fission should also be investigated.
Finally, the influence of fragment shell effects over the shape of the  
asymmetric scission line itself should also be considered.

\backmatter

\bmhead{Acknowledgements}

R. N. B. would like to thank Luis M. Robledo for providing the
\textsc{HFBaxial} code.
This work has been supported by the Australian Research Council 
under Grant No. DP190100256.
Some calculations were performed using computational resources provided by the Australian Government through the National Computational Infrastructure (NCI) under the ANU Merit Allocation Scheme.
N.-W. T. L. acknowledges the support of the Australian Commonwealth through the Australian Government Research Training Program (AGRTP) Scholarship.

\begin{appendices}

\section{\label{SPEannex}$^{180}$Hg and $^{90}$Zr Single Particle Energies}

\begin{figure*}
  \centering 
   \includegraphics[width=0.7\linewidth ]{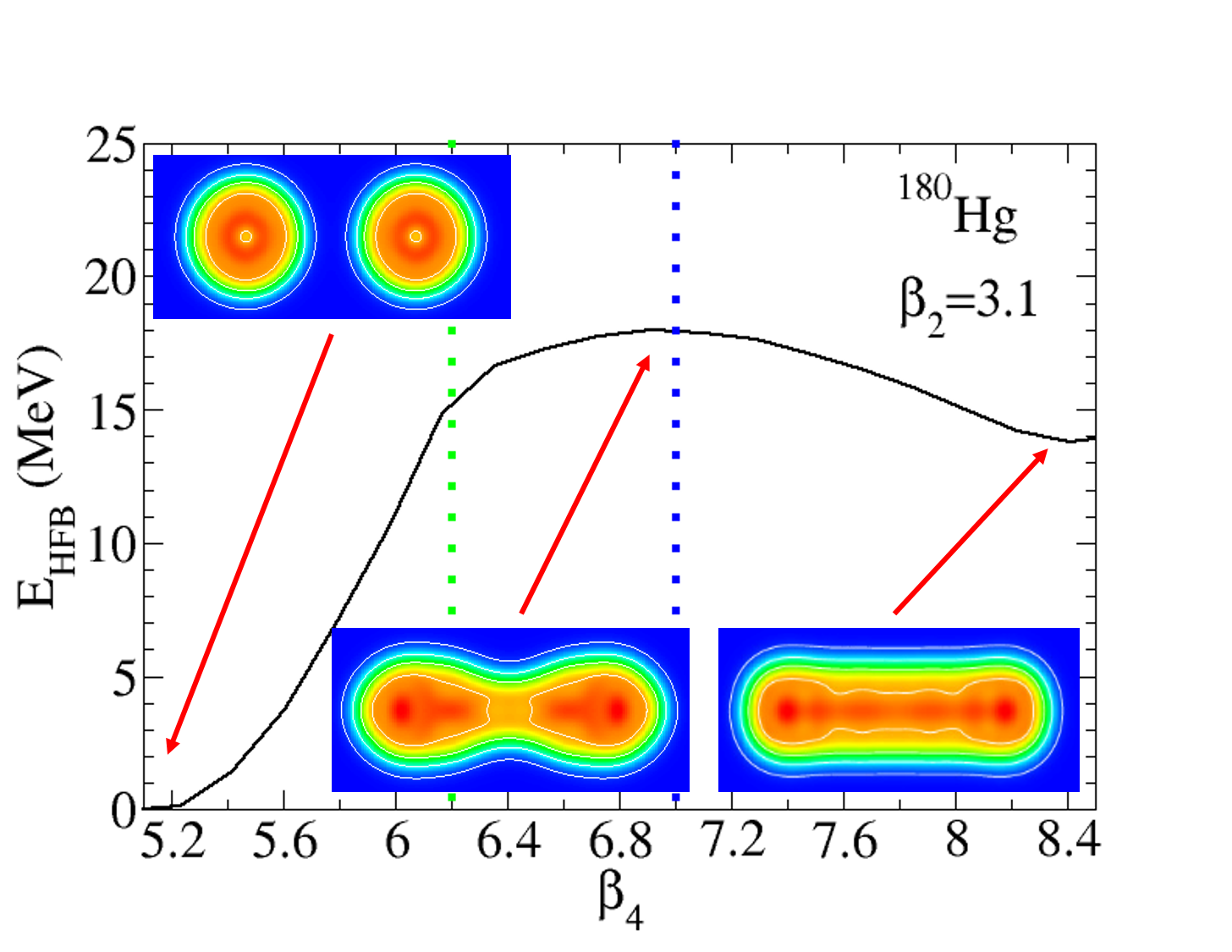}
\caption{Relative HFB energy in the $\beta_2=3.1$ 
slice from the $^{180}$Hg symmetric PES in Fig.~\ref{fig::PESsym} as a function of $\beta_4$ from well separated fragments in the fusion valley to the fission valley energy minimum.
Green and blue dot lines indicate the $\beta_4$ scission and top of the ridge configurations, respectively.}
\label{fig::slice100b} 
\end{figure*}

This Appendix presents the transition between two well
separated $^{90}$Zr fragments to the compound $^{180}$Hg nucleus
in a transverse slice of the symmetric valley ($\beta_3$ imposed to be zero).
In Fig.~\ref{fig::slice100b} the HFB energy of the whole system is plotted
with respect to $\beta_4$ from the bottom of the fusion valley to
the bottom of the fission one at a fixed $\beta_2=3.1$.
This latter value is the elongation of the final
point of the transitional asymmetric to symmetric valley depicted in
Fig.~\ref{fig::PESgap}. 
The barrier height going from the bottom of the fission valley to the fusion
one is about $4$~MeV in this region, preventing $^{180}$Hg from scissioning
at intermediate elongations.
Looking at elongations greater than $\beta_2 = 3.1$ the energy barrier between the fusion valley and the fission one decreases
while the symmetric valley rises
in energy. Finally the energy barrier vanishes
at $\beta_2 \sim 4.5$ and the symmetric valley ends up at an energy higher than the asymmetric valley, prohibiting any elongated symmetric mode from appearing in the
low energy fission of $^{180}$Hg.

Starting from the fusion valley with two
spherical $^{90}$Zr fragments (see left inner panel of Fig.~\ref{fig::slice100b}) the system progressively
passes through the scission point (green dotted line) and
evolves to become a dinuclear system composed of two deformed $^{90}$Zr
prefragments up to the saddle point (see right inner panel) denoted by
the blue dotted line. At higher $\beta_4$ deformation the neck disappears
and $^{180}$Hg loses its dinuclear character at the bottom of the fission
valley at $\beta_{4}=8.4$ (see inner panel of Fig.~\ref{fig::PESsym}(a)).

\begin{figure*}
  \centering 
   \includegraphics[width=0.49\linewidth ]{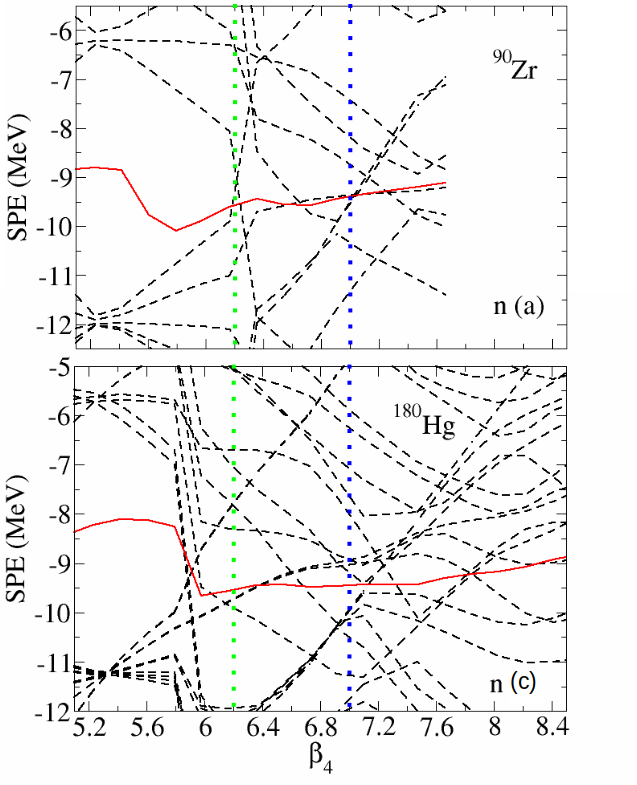}
   \includegraphics[width=0.49\linewidth ]{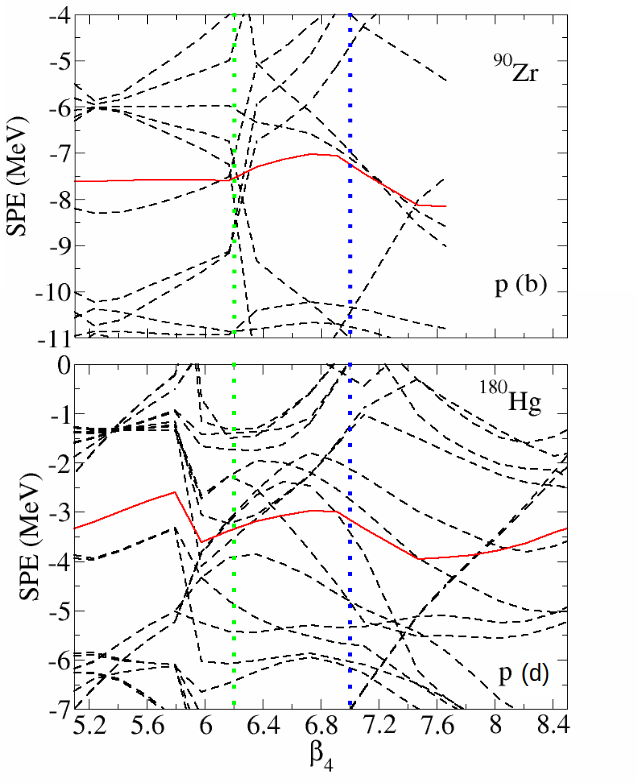} \\
\caption{Neutron (c,a) and proton (d,b) single particle energies of $^{180}$Hg and its symmetric fragment $^{90}$Z along the symmetric $\beta_2=3.1$ 
slice of Fig.~\ref{fig::slice100b}, respectively. For each $\beta_4$ deformation, $^{180}$Hg is separated into two $^{90}$Zr symmetric fragments, whose deformations are extracted from the $^{180}$Hg spatial density. Then HFB calculations are performed on the $^{90}$Zr configurations to determine the $^{90}$Zr single particle energies.}
\label{fig::SPE_Hg_Zr} 
\end{figure*}

In Fig.~\ref{fig::SPE_Hg_Zr} the neutron and proton 
single particle energies around the Fermi levels of
$^{180}$Hg are plotted, derived from the mean field states in Fig.~\ref{fig::slice100b} and the
$^{90}$Zr ones obtained  by separating the fragments for $^{180}$Hg configurations at small $\beta_4$ deformation ($\beta_4 \le 7.7$). 
To do so the multipole moments up to $\beta_6$ are calculated for each
$^{180}$Hg configuration and are then used as constraints for $^{90}$Zr accordingly to the method depicted in \ref{sec::framework}.
Scission and saddle points are depicted by green and blue dotted
lines respectively as in Fig.~\ref{fig::slice100b}.
At small $\beta_4$ deformations all the single particle states
of $^{180}$Hg are nearly degenerate as the whole system is built
from two well separated spherical $^{90}$Zr fragments. 
When increasing the $\beta_4$
deformation the compound nucleus experiences some strong variations in 
the single particle energy picture as a nascent neck links the two fragments.
Note that these variations are signatures of discontinuities in the neck operator variable.
They can be solved by linearly decreasing the neck value at the discontinuity making the transition
from the fission valley to the fusion one smoothed. Smoothing procedures as depicted in Ref.~\cite{LauPRC22,lasseri2024,Carpentier2024} 
could also be applied to get a more physical transition.
$^{90}$Zr also experiences strong structure variations around the scission
point since its own octupole deformation starts increasing.
At the saddle point the $^{90}$Zr deformed prefragment displays some neutron 
single particle energy intruders rising and some proton ones decreasing
in energy. These intruders may be identified among the $^{180}$Hg level densities shown in Fig.~\ref{fig::SPE_Hg_Zr},
reflecting the local change of structure at the ridge.
Beyond this deformation it is harder and harder to correlate $^{180}$Hg single
particle energies with $^{90}$Zr ones.

\end{appendices}


\bibliographystyle{apsrev4-2}
\bibliography{Gen_Biblio}

\end{document}